\documentclass[twocolumn,secnumarabic,superscriptaddress,amssymb, nobibnotes, aps, prb,floatfix]{revtex4-2} %fleqn: make all eqns left alighed  ,showpacs
\usepackage{epsfig,amssymb,amsmath,graphicx,subfigure,hyperref  }  % pdfpages and graphicx are added   hyperref(set hyperlink for content)
\usepackage{mathrsfs}
\usepackage{color}
\usepackage{multirow}
\usepackage{tabularx}
\usepackage[normalem]{ulem}
\usepackage{physics}

\renewcommand{\vec}[1]{\boldsymbol{#1}}
\usepackage{xcolor}

\newcommand{\Chir}{\chi}
\newcommand{\jth}{p_\theta}
\newcommand{\spin}[1]{|#1\rangle}
\begin{document}

\title{Rotating Majorana Zero Modes in a disk geometry}
\author{Liu Yang} 
\email{liu.yang-2@manchester.ac.uk}
\affiliation{Department of Physics and Astronomy, The University of Manchester, Manchester M13 9PL, United Kingdom }
\author{Alessandro Principi}
\affiliation{Department of Physics and Astronomy, The University of Manchester, Manchester M13 9PL, United Kingdom }
\author{Niels R. Walet}
\affiliation{Department of Physics and Astronomy, The University of Manchester, Manchester M13 9PL, United Kingdom } % \altaffiliation[Also at ]
%\author{}
%\affiliation{}

\begin{abstract}
We study the manipulation of Majorana zero modes in a thin disk made from a $p$-wave superconductor in order to understand their use as a building block for topological quantum computers. We analyze the second-order topological corner modes that arise when an in-plane magnetic field is applied and calculate their dynamical evolution when rotating the magnetic field, with special emphasis on non-adiabatic effects. We characterize the phase transition between high-frequency and near-adiabatic evolution using Floquet analysis. We show that oscillations persist even in the adiabatic phase because of a frequency-independent coupling between zero modes and excited states, which we have quantified numerically and analytically. These results show that controlling the rotation frequency can be a simple method to avoid the non-adiabatic errors originated from this coupling and thus increase the robustness of topological quantum computation.

\end{abstract}
\maketitle

\section{Introduction}
Topological quantum computing is a promising approach to quantum information processing since it provides robustness against local perturbations, see for instance \cite{NonAbelianRMP,topoquantum2018}. One way to implement topological quantum computing is through manipulating Majorana zero modes. Using their non-Abelian statistics, one can use adiabatic exchange of Majorana zero modes to construct topological quantum gates. One of the attractive aspects of topology is the robustness against local perturbations with symmetry protection, and thus if one can produce Majorana zero modes, they may provide a very robust platform for quantum computation \cite{Kitaev_2001,Nabel_2001,Milestone_2016,topoquantum2018}.

One particular set of systems of interest consists of time-reversal invariant topological superconductors (TSCs) and superfluids in two and three dimensions. These can host gapless counter-propagating Majorana edge modes \cite{Qi_2009}. One of the candidates for the realization of $p$-wave TSCs are doped Bi$_2$Se$_3$ compounds \cite{Bi2Te3_2009,liang_2010,Hamiltonian2010,liang_2012,oddparity2order2016,phongPRB,RSB2017}, and several experiments have indeed given evidence for $p$-wave pairing in Cu$_x$Bi$_2$Se$_3$ \cite{experiment_2010,experiment_2011}. 

Recent studies show that when a perturbation that breaks the symmetry protecting the topology is applied, the resulting system can host lower-dimensional topologically protected gap-less states called hinge or corner modes \cite{higher_order_science,Majorana_corner,HOSC_odd}. Such Majorana corner modes have been considered in various 2D systems \cite{Kane_Rashba,Majorana_corner,highT_corner2018yan,highT_corner2018zhang,phongPRB,s_corner2020}. 

In our work consider a quasi-two-dimensional system consisting of a thin disk of doped Bi$_2$Se$_3$, described by a model developed by Phong \emph{et al.} \cite{phongPRB}, which can be turned into a second-order TSC. In that approach, an in-plane magnetic field breaks the time-reversal symmetry and stabilizes Majorana zero modes on the edge of the system. Adiabatically rotating the orientation of the magnetic field results in the movement of Majorana zero modes along the edge of the disk. This fact can be used to define a protocol for braiding Majorana zero modes of multiple disks, clearly showing the advantage of working with a disk geometry\cite{CorbinoGeometry,RotatingMajorana2020,Songbo2020,SongboPRR}.

Realistic braiding for topological quantum computation cannot be perfectly adiabatic. Thus, it is necessary to understand the nonadiabatic corrections that affect the movement of the Majorana zero modes \cite{Nonadiabatic2011,Nonadiabatic2013,Boosting,dynamic_wire2015,optimal2015,Error2019}. In this paper, we will analyze in detail the evolution of Majorana states in the model considered in Ref.~\cite{phongPRB}. As discussed in section \ref{sec:model}, we derive a new effective one-dimensional description of the edge states, which allows us to analyze the behavior of the Majorana zero modes in detail. Numerical results for a fixed orientation of the magnetic field are shown in Sec.~\ref{sec:NumEff}. We then turn our attention to the dynamical evolution of zero modes for a rotating field as a function of frequency in Sec.~\ref{sec:Evo}. We concentrate on a detailed analysis of the overlap between the final state and the initial Majorana state in a one-cycle rotation. Through Floquet analysis of the evolution operator \cite{floquet1965,Floquet1977}, we find that, at the transition point between the high-frequency sudden phase and the low-frequency near-adiabatic phase, the quasi-energy spectral gap closes. 
It is shown that the approach to the adiabatic limit is not uniform, but we find pronounced oscillations of the zero mode probability amplitude, which can be described by a frequency-independent tunneling in a co-rotating frame. This oscillatory behavior is problematic, since it makes it harder to control quasi-particle poisoning. In Sec.~\ref{sec:nona}, we calculate the oscillation frequency using perturbation theory. We gain a better analytic understanding of the source of these problems. Furthermore, we identify field-rotation frequencies that are best for high-fidelity braiding of Majorana zero modes. Finally, we draw some conclusions in Sec.~\ref{sec:conc}.

\section{Model}\label{sec:model}
We study a minimal, universal model for a time-reversal-invariant two-dimensional $p$-wave TSC.
This model occurs in various contexts, one of which is the description of a thin slab of three-dimensional triplet time-reversal superconducting phase in a topological superconductor such as doped Bi$_2$Se$_3$~\cite{Bi2Te3_2009,liang_2010,Hamiltonian2010,liang_2012,oddparity2order2016,phongPRB,RSB2017}.  In our work, following~\cite{phongPRB} we consider this 2D topological superconductor system with a circular geometry, i.e., as a disk of radius $R$. Making the assumption that the Fermi energy lies well inside the conduction band, we can write the state of the system as a four-component wave function, $\Psi(\vec{r})=\left(\begin{array}{cccc}{\Psi_{\downarrow}^e(\vec{r})} & {\Psi_{\uparrow}^h(\vec{r})} & {-\Psi^h_{\downarrow}(\vec{r})} & {\Psi_{\uparrow}^e(\vec{r})}\end{array}\right)^{T}$,
as shown in Ref.~\cite{phongPRB}. Here the subscripts $\uparrow$ and $\downarrow$ label the spin of the electron or hole, respectively. The superscripts ``$e$'' and ``$h$'' denote the components in Nambu (particle-hole) space.

We can then express the Bogoliubov–de Gennes Hamiltonian in polar coordinates as
\begin{align}
&H(\vec{r})=\bar{\epsilon}_{F}\left(\frac{1}{k_F^2 r} \frac{\partial}{\partial r}r \frac{\partial }{\partial r}+\frac{1}{k_F^2 r^{2}} \frac{\partial^{2} }{\partial \theta^{2}}-1\right) \tau_{z}\otimes s_z\nonumber\\&+\frac{\bar{\epsilon}_{F}\sqrt{2\gamma}}{k_F} \left(\begin{array}{cc}{0} & {-e^{\mathrm i\theta}(\frac{\partial}{\partial r}+\frac{\mathrm{i}}{r}\frac{\partial}{\partial\theta})} \\ {e^{-\mathrm{i}\theta}(\frac{\partial}{\partial r}-\frac{\mathrm{i}}{r}\frac{\partial}{\partial\theta})}  & {0}\end{array}\right)\otimes s_0\,.  \label{eq:first}
\end{align}
In this equation, the Pauli matrices $\tau_i$ act in Nambu space, i.e., on the particle-  and hole components of the wave function, and $s_i$ are the spin matrices. The parameters $\bar{\epsilon}_{F}$ is the difference between the Fermi energy and band gap of the original three-dimensional insulator ($E_G$), while the Fermi wavenumber 
$k_F = \sqrt{E_G\bar{\epsilon}_{F}}/(\hbar v_F)$. Both the gap energy $E_G$ and Fermi velocity $v_F$ should be determined from experimental data. The dimensionless parameter $\gamma$ is defined by the measured superconducting gap $\Delta_\text{exp}$ and $\bar{\epsilon}_F$ as $\Delta_\text{exp}^2/(2\bar{\epsilon}_{F}^2)$ \cite{phongPRB}. Here, we select a gauge where the superconducting order parameter is real. We will use $\hbar=1$ and shall work at zero temperature throughout the rest of the paper. The TSC model \eqref{eq:first} possesses time-reversal and particle-hole symmetry. A detailed discussion of the symmetry operators, including the time reversal $\hat{T}$ and charge conjugation $\hat{C}$, can be found in Appendix \ref{app:wfsym}.

The spectrum of 2D $p_x\pm\mathrm{i}p_y$ superconductors is gapped, but in finite systems there are in-gap edge states which are protected by the topological properties of the bulk. In such systems, the topological phase of the bulk is characterized in terms of the first Chern number, while its sign determines the propagation direction (chirality) of edge states \cite{TKNN,WKB_disk,Read_pairing,Qi_2009,phongPRB,Sato_2017,Bernevig}. In the time-reversal invariant system considered here, the subspace with left-handed chirality has Chern number $+1$, and the other subspace has  Chern number $-1$. Edge states are not only characterized by chirality, but also by their total angular momentum, with the eigenvalues given by a half-integer angular momentum quantum number $j=\ell+1/2$ ($\ell$ is an integer). The total angular momentum operator for a chiral superconductor is $L_z=-\mathrm{i}\partial_{\theta}-\tau_z/2$ \cite{phongPRB}, thus we can obtain the corresponding representation for a time-reversal-symmetric superconductor by combining two copies of $L_z$
\begin{equation}
    J_{z}=\left(-\mathrm{i}\partial_{\theta}-\frac{\tau_z}{2}\right)\otimes s_{0}\,. \label{eq:Jz}
\end{equation} 
When $\gamma$ is small and $R$ is large, the radial wave functions of the low-energy edge states satisfying hard-wall boundary conditions $\Psi(r=R,\theta)=0$ are approximately independent of $j$. 

In terms of a dimensionless radius and Fermi momentum,  $\rho=r/R$ and $\lambda=k_FR$, we can approximate the wave function for $\lambda\gg1$, 
\begin{align}
\Psi_j^\pm(\rho,\theta)&\approx\frac{f(\rho)}{\sqrt{4\pi}}e^{\mathrm{i}j\theta}\left(\begin{array}{c}e^{\mathrm{i}\theta/2}  \\
\mp e^{-\mathrm{i} \theta/2}
\end{array}\right)\otimes\spin\pm,\label{psi_j}\\
f(\rho)&=-\frac{\widetilde{\mathcal{N}}}{\sqrt{\rho}}e^{\lambda\sin{\xi}\rho}\sin{[\lambda\cos{\xi}(1-\rho)]}\label{radial_f}.
\end{align}
where we define the two-component spinors $\spin+=(1,0)^T$  and $\spin-=(0,1)^T$ to correspond to the left-hand and the right-hand chiral state, respectively.  
The function $f(\rho)$ is the radial component of the wave function and is independent of $j$, and is normalized as
\begin{equation}
    \int_0^1 f^2(\rho)\rho \mathrm{d}\rho=1
    \,.
\end{equation}
Finally, the parameter $\xi$ in Eq.~(\ref{radial_f}) is defined as
\begin{equation}
    \xi=\arctan{[\sqrt{2\gamma}/(1-\gamma)]}/2\,.
\end{equation}
The detailed derivation of the approximate wave function is given in Appendix \ref{app:wfsym}. We can make a rough estimate of the parameters in the wave function, and we see that the radial wavenumber is approximated as $k_F$ and the radial decay length of the wave function from the boundary to the interior of the disk is $k_F\sin{\xi}\approx \Delta_{\text{exp}}/(2k_F\bar{\epsilon}_F)$ which should be much smaller than the disk’s radius $R$ for this model to be valid. In fact, the angular momentum of the edge states must be restricted to the region $|j| <\lambda\sqrt{1-\gamma/2}$. For larger $j$, such states lie outside the superconducting gap and are embedded in the continuum, and are therefore delocalized~\cite{phongPRB}. We have checked the validity of this inequality in a tight-binding model, as discussed in Appendix \ref{app:TB}. Note that, as shown in Ref.~\onlinecite{Locking_2012,phongPRB}, the scalar disorder does not mix in-gap states.

\begin{figure}
\centering
\includegraphics[width=\columnwidth]{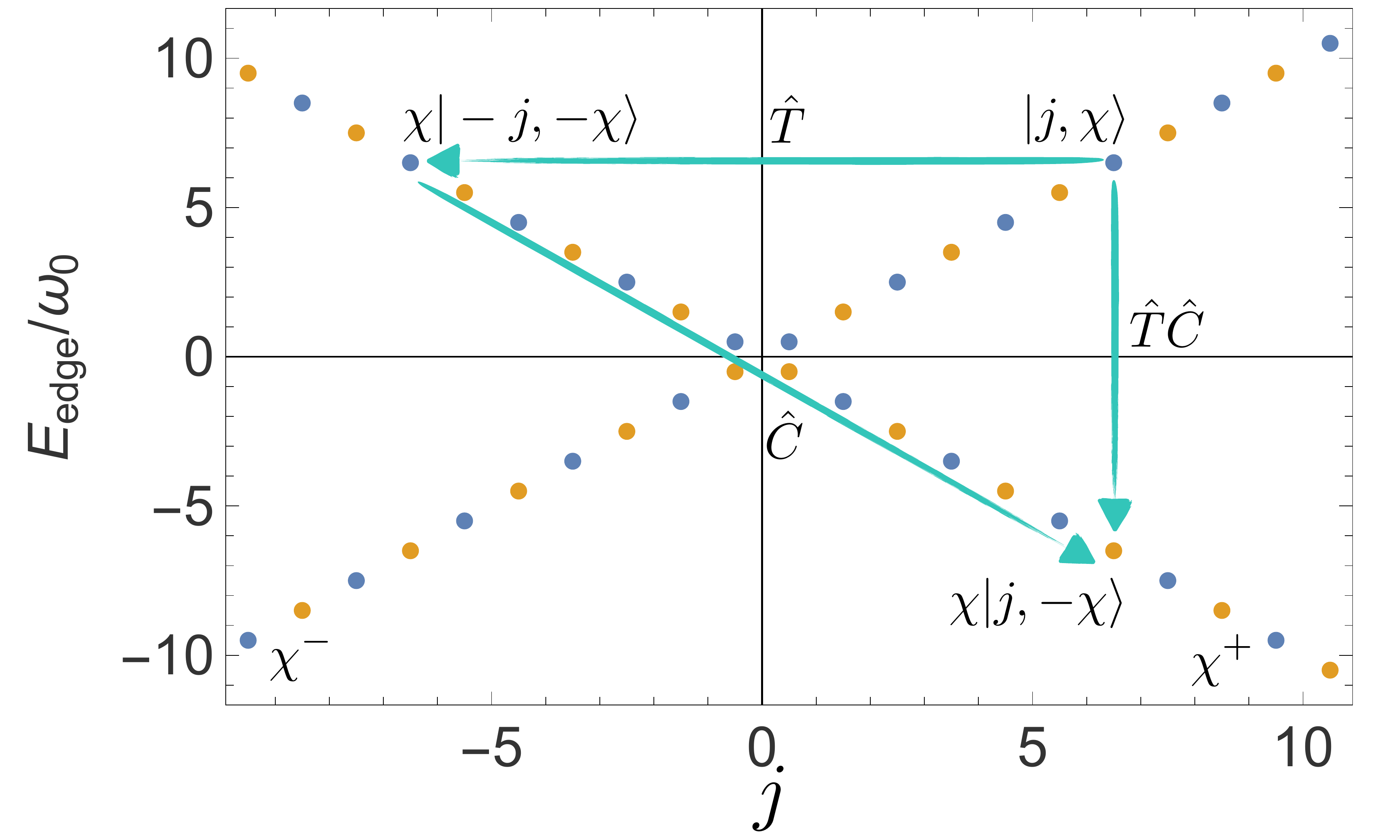}
\caption{Schematic representation of the energies and symmetry transformations of edge states. The operator $\hat{T}$ is the time reversal operation, while $\hat{C}$ is charge conjugation. $\hat{T}$ connects a state to its time-reversed partner, which has opposite angular momentum and chirality, while $\hat{C}$ only reverses $j$, leaving chirality intact. The blue points denote edge states belonging to the set $\mathbf{J}_1$ while yellow points are members of set $\mathbf{J}_2$, both defined in Eq.~(\ref{eq:J1J2}).}\label{twobranch}
\end{figure}

In the limit of large $R$, the energy of $\Psi_j^\pm(\rho,\theta)$ is\cite{WKB_disk}
\begin{align}
E_j^\pm=\mp \omega_0 j, \label{eq:Ejpm}
\end{align}
where \begin{align}\omega_0=\bar{\epsilon}_{F}\frac{ \sqrt{2 \gamma}}{\lambda}=\frac{\Delta_{\text{exp}}}{k_FR}\label{w0}.\end{align}
That is to say, the energy of edge states is approximately proportional to their angular momentum quantum number $j$. In what follows, a chiral edge state with total-angular-momentum quantum number $j$ is denoted as $|j,\Chir \rangle$, where $\Chir=\pm$ denotes the chirality, clockwise ($+$) and anti-clockwise ($-$), respectively. The two sets of chiral edge states sets are denoted as $\Chir ^+$ and $\Chir ^-$ respectively. We define a chirality operator $\hat{X}$ with eigenvalues $\pm1$, represented by
\begin{align}
X=\tau_0\otimes s_z,
\end{align}
which satisfies $\hat{X}|j,\Chir \rangle=\Chir|j,\Chir \rangle$. The edge-state energies and symmetry transformations among these states are shown schematically in Fig.~\ref{twobranch}.

Finally, Majorana zero modes are generated by breaking time-reversal symmetry~\cite{phongPRB,Majorana_corner}. This can be achieved by applying an \textit{in-plane} magnetic field, which turns the system into a higher-order TSC \cite{higher_order_science,Majorana_corner}.
The Zeeman Hamiltonian describing the coupling to an external magnetic field is 
\begin{align}
H_{Z}(\phi)=E_{Z}\left(\begin{array}{cc}
0 & e^{\mathrm{i} \phi} \\
e^{-\mathrm{i} \phi} & 0
\end{array}\right)\otimes s_{x},\label{H_Zeeman}
\end{align}
where $E_{Z}=\mu_{e} B_{Z}$ is the Zeeman energy of the field, $\mu_{e}$ is the magnetic moment of the electron, $B_{Z}$ is the field strength, and $\phi$ is the angle of that field with the $z$ axis. Note that we work in the regime where the Zeeman field is weak so as not to spoil the superconducting phase~\cite{H_limit}. 

While the in-plane Zeeman field breaks the time-reversal-symmetry, the Hamiltonian still retains the particle-hole symmetry and invariance under the product of time-reversal and chirality:
\begin{align}
\{\hat{H}+\hat{H}_Z,\hat{C}\}&=0,\label{eq:symP}\\
[\hat{H}+\hat{H}_Z,\hat{T}\hat{X}]&=0.\label{eq:symTC}
\end{align}

Since the in-plane Zeeman field couples states of opposite spin, it couples an edge state with angular momentum $j$ and chirality $\Chir$ to the $j\pm1$ states with the opposite chirality $-\Chir$. This coupling scheme leads to two uncoupled subspaces spanned by two sets of edge states which are related by angular-momentum inversion:
\begin{align}
    \mathbf{J}_1 &=(\ket{j_{\max },-},\ket{j_{\max }-1,+},\ldots,\ket{-j_{\max }+1,-})\,,\nonumber\\ \mathbf{J}_2&=(\ket{-j_{\max },-},\ket{-j_{\max }+1,+},\ldots,\ket{j_{\max }-1,-})\label{eq:J1J2}.
\end{align}
 Here $|j|$ must be constrained to $|j| <j_\text{max}=\lambda\sqrt{1-\gamma/2}$ \cite{phongPRB}. 

Since the edge states are concentrated at the boundary, a complete two-dimensional treatment is unnecessarily complex: we would like to construct a one-dimensional effective Hamiltonian that only involves said edge states. Note that there is only one term in Eq.~(\ref{psi_j}) dependent on $j$ which is $e^{\mathrm{i}j\theta}$. We can thus write the effective Hamiltonian ignoring the radial part of the wave function, by considering only the angular part $e^{\mathrm{i}j\theta}\spin\pm/\sqrt{2\pi}$. In this basis, the operator $J_z$ defined in Eq.~\eqref{eq:Jz}, maps to $p_\theta=-\mathrm{i}\partial_\theta$. This reduces the two-dimensional model on a disk to an effective one-dimensional model on a ring. This is valid as long as the decay length of edge states is much smaller than the disk's radius. The reduced edge Hamiltonian takes the form
\begin{align}
H_\text{edge}(\theta;\phi) &=\omega_0\left[-\jth  \sigma_{z}-2 \epsilon \sin (\theta-\phi) \sigma_{y}\right].\label{Hphi}
\end{align}
In the above equation, $\epsilon=E_{Z}/(2\omega_0)$ is the dimensionless magnitude of the Zeeman field, and this Zeeman term is a Dirac mass term \cite{Bernevig}. It is well known that a change of the sign of this term can create domain-wall soliton states\cite{soliton,domain_wall}.

The reduced wave function can be written as \begin{align}\psi(\theta)=
\begin{pmatrix}\psi^+(\theta), \psi^-(\theta)\end{pmatrix}^T,\end{align} which must obey anti-periodic boundary condition because the angular momentum quantum number $j$ is half-integer. The two components of this spinor correspond to the angular functions for each of the two chiral blocks $\Chir ^\pm$.

In this effective model, the three basic symmetry operators of the original 2D model, the chirality operator $\hat{X}$, charge conjugation $\hat{C}$ and time reversal $\hat{T}$ are now represented by the operators $\sigma_z$, $\mathcal{K}$ and $-\mathrm{i}\sigma_y\mathcal{K}$, respectively. Here $\mathcal{K}$ is the complex conjugation operator.

\section{Numerical and Effective time-independent solutions} \label{sec:NumEff}
\begin{figure*}[tbp]
\centering
\includegraphics[width=0.8\textwidth]{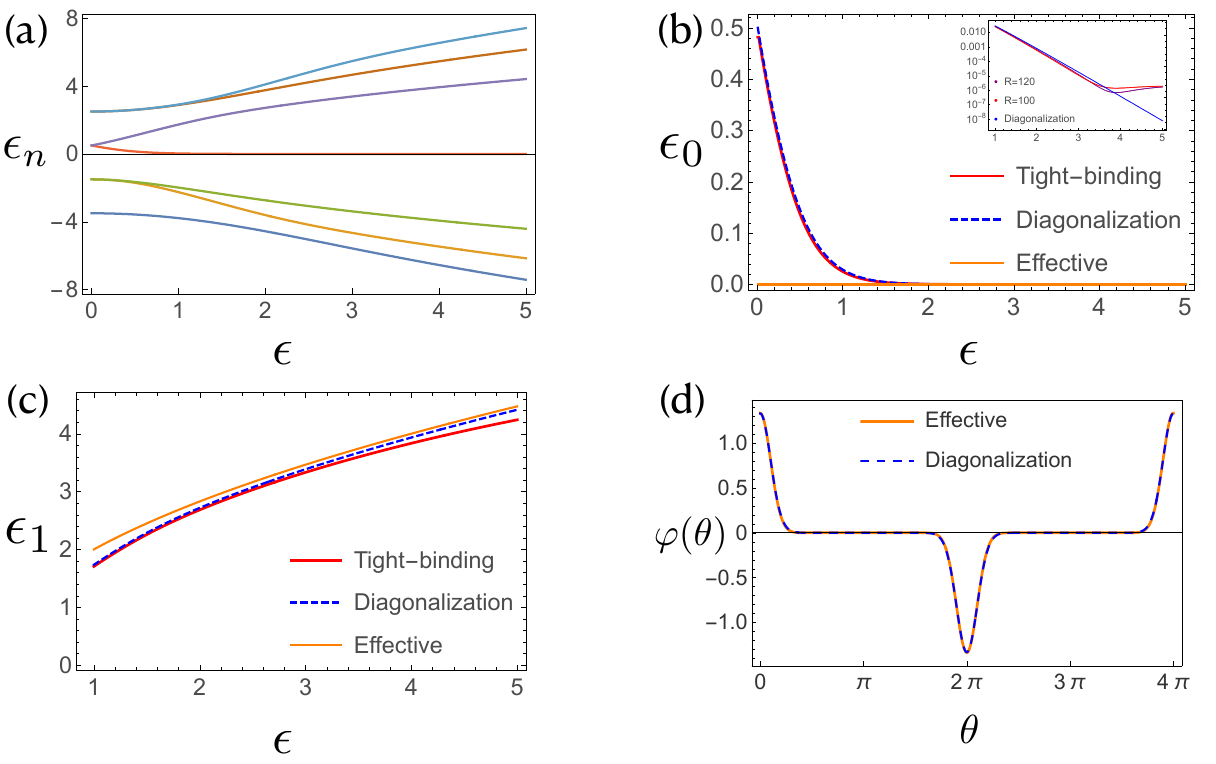}
\caption{(a)-(c): Low-energy eigen-energies of the edge space $\mathbf{J_1}$ (defined in Eq.~(\ref{eq:J1J2})) under the influence of an in-plane magnetic field as a function of the dimensionless Zeeman energy $\epsilon=E_Z/(2\omega_0)$. (a) Shows the seven smallest eigenvalues of $h(\theta)$ (defined in Eq.~(\ref{htheta})), obtained by a numerical diagonalization method within the $\mathbf{J}_1$ subspace. We can see that a zero mode can be identified when $\epsilon$ exceeds 2. In (b) and (c), we compare the ground state and first-excited state energies, respectively, obtained by the numerical matrix diagonalization (blue dashed line), tight-binding model (red line--for details see Appendix \ref{app:TB}) and the effective results (orange line--for details see Eq.~(\ref{dirac_sqrt})). We can see that the results of the three methods show excellent agreement in the region we focus on ($\epsilon>1$). In the insert in figure (b), we further compare the tight-binding and the diagonalization and show that they exhibit the same exponential suppression of the zero-mode energy to a value of $10^{-6}$ at $\epsilon=3.5$, dependent on the lattice pacing in our tight-binding model. (d): The angular function $\varphi(\theta)$ of the Majorana zero mode (defined in Eq.~(\ref{angular_phi})) when $\epsilon=5$ obtained using a numerically exact diagonalization and the effective Hamiltonian. We see that the effective expression is practically indistinguishable from the numerical result obtained by matrix diagonalization.}\label{hsolution}
\end{figure*}
We now analyze the effective Hamiltonian $H_\text{edge}$. For simplicity, we divide it by the frequency $\omega_0$, and due to rotational symmetry only consider $\phi=0$,
\begin{align}
        h(\theta)=H_\text{edge}(\theta;0)/\omega_0\,.\label{htheta}
\end{align}
First of all, we expand the Hamiltonian in the eigenstates of $\jth $. Using particle-hole symmetry, the solutions in the two subspaces  $\mathbf{J}_1$ and $\mathbf{J}_2$, Eq.~\eqref{eq:J1J2}, can be mapped onto each other. Hence, we can focus our discussion on the subset $\mathbf{J}_1 $ and we take $j_{\max}=48+1/2$ as a reasonable illustration. By numerical matrix diagonalization, we find the full spectrum of the effective edge Hamiltonian. We show the energy of the seven states nearest zero as a function of the scaled Zeeman energy $\epsilon$ in Fig.~\ref{hsolution}(a) and compare the solution of the effective edge Hamiltonian with a more complete solution using a tight-binding method (see Appendix \ref{app:TB} for more details) in Figs.~\ref{hsolution}(b) and (c). In Figs.~\ref{hsolution}(a) and (b), we can see that a zero mode appears as $\epsilon>2$. By analyzing the results of the diagonalization of the reduced Hamiltonian in the $\mathbf{J}_1$ subspace, we find that zero-energy states decay faster than the best exponential fit $\epsilon_0\approx0.528\exp{(-2.66\epsilon)}$, where $\epsilon=0.5E_Zk_FR/\Delta_{\exp}$. Thus, the energy splitting of zero modes can be suppressed exponentially by increasing the radius $R$ and the Zeeman energy $E_Z=\mu_eB_Z$ (defined in Eq.~(\ref{H_Zeeman}). In Fig.~\ref{hsolution}(b) and (c), we can see a good agreement between the three methods. We should mention that when $\epsilon$ becomes large the results of the numerical diagonalization and effective methods slightly deviate from the tight-binding model since it contains more states of the 2D bulk. This difference is small in the region of interest and the zero-mode energy is exponentially suppressed to $10^{-6}$ at  $\epsilon=3.5$ (see the insert in Fig.~\ref{hsolution}(b)). As we increase the number of lattice points of the tight-binding calculation, the exponential suppression persists over a larger interval, and thus the flattening of the suppression is due to lattice effects.

We write the $\theta$ representation of the two-component wave function of the zero mode in the $\mathbf{J}_1$ subspace defined in Eq.~\eqref{eq:J1J2}, as 
\begin{equation}
\psi_0(\theta)=(\psi^+_0(\theta),\psi^-_0(\theta))^T\,.
\end{equation}
The general expansion of $\psi_0(\theta)$ is shown in Eq.~(\ref{psi0_sum}). The zero mode in the other ($\mathbf{J}_2$) subspace has wave function $\psi_0^\ast(\theta)$. Using the symmetry operations in Eqs.~(\ref{eq:symP}) and (\ref{eq:symTC}), we can prove that 
\begin{equation}
\hat{T}\hat{X} \psi_0(\theta)=\sigma_x\mathcal{K}\psi_0(\theta)=\psi_0(\theta)\,.\label{psi_symmetry}
\end{equation}
We can thus superimpose the degenerate zero-energy states and express the wave functions of the two Majorana zero modes as
\begin{align} 
\psi^m_{1}(\theta)&=\frac{\psi_{0}(\theta)+\psi^\ast_{0}(\theta)}{\sqrt{2}}=\varphi(\theta)\,\frac{\spin++\spin-}{\sqrt{2}},\label{eq:main_psi_1_m}\\
\psi^m_{2}(\theta)&=-\frac{\mathrm{i}[\psi_{0}(\theta)-\psi^\ast_{0}(\theta)}{\sqrt{2}}=\varphi(\theta-\pi)\,\frac{\spin+-\spin-}{\sqrt{2}},\label{eq:main_psi_2_m}
\end{align}
where 
\begin{align} \varphi(\theta)&=\psi^+_0(\theta)+\psi^-_0(\theta)=2\text{Re}\{\psi^+_0(\theta)\}.\label{angular_phi}\end{align} 
We give the full expressions of the wave functions $\psi^m_{1}(\theta)$ and $\psi^m_{2}(\theta)$, as well as their inter-relations in Appendix \ref{app:onedangle}.

We show the angular function $\varphi(\theta)$ calculated through matrix diagonalization for the region $\theta\in[0,4\pi]$ in Fig.~\ref{hsolution}(d). It is clear from that figure that the Majorana zero mode $\psi^m_{1}$ is localized around the point $\theta=0$, and similarly $\psi^m_{2}$ is localized around $\theta=\pi$.

After analyzing the numerical results, we would like to get an approximate analytical description for further analysis.
Due to the antiperiodic boundary conditions, we cannot approximate the Hamiltonian $h(\theta)$ by linearizing the Dirac mass term around the point it changes sign ($\theta=0$ or $\theta=\pi$) and solve the two independent Dirac-Landau level spectra as in Refs.~\cite{CorbinoGeometry,RotatingMajorana2020}. 
The method used in \cite{phongPRB} mixes the two uncoupled sets and thus only describes half the spectrum.
To avoid these problems, we will work in the $j$-representation of the edge Hamiltonian but individually solve for a single set, which we take to be $\mathbf{J}_1$.
To achieve this goal, we first relabel the angular momentum states as 
\begin{align}|k;\pm\rangle=\,|j=2k\mp 1/2,\pm\rangle
\end{align} 
to denote the basis in $\mathbf{J}_1 $ with integer $k$, and then make the approximation that the low-energy wave functions vary slowly in $k$ so that we can approximately treat $k$ as continuous in this region. Hence, we express the Hamiltonian in terms of $k$ and derivatives with respect to $k$,
\begin{align}
h(k)&=\left(\begin{array}{cc}{-2k+1/2} & {\epsilon(e^{-\partial_k}-1)} \\ {\epsilon(e^{\partial_k}-1)}&{2k+1/2} \end{array}\right).\label{k_rep}
\end{align}
Note that $h(\theta)$ defined in Eq.~(\ref{htheta}) differs from $h(k)$ in the Eq.~(\ref{k_rep}); the latter is a continuous Fourier representation for $h(\theta)$ in the subspace $\mathbf{J}_1$ only. To solve most compactly for the eigenstates of $h(k)$, we first apply a unitary transformation $Q^\dag h(k)Q$ to $h(k)$, with  
\begin{align}Q=\frac{1}{\sqrt{2}}\left(\begin{array}{cc}{e^{-\partial_k/4}} & {-e^{-\partial_k/4}} \\ {e^{\partial_k/4}}&{e^{\partial_k/4}} \end{array}\right).\label{eq:defQ}\end{align}
This substantially simplifies the matrix structure of $h(k)$ since it sets the diagonal elements to zero. We then expand the result to the first order in the derivative $\partial_k$ to obtain an effective Hamiltonian. The latter can then be simply expressed in terms of harmonic oscillator creation and annihilation operators, $a^\dagger$ and $a$ associated with the variable $k$ \cite{landau,Sakurai}, as
\begin{align}
h_{\text{eff}}(k)&=Q^\dag h(k)Q=2\sqrt{\epsilon}\left(\begin{array}{cc}{0} & {a^\dag} \\ {a}&{0} \end{array}\right),\\
a&=\frac{k}{\sqrt{\epsilon}}+\frac{\sqrt{\epsilon}}{2}\partial_k.
\end{align}
The spectrum of the effective Hamiltonian is \cite{Boosting}
\begin{align}\epsilon_n=\text{sign}(n)\sqrt{4\epsilon|n|}\label{dirac_sqrt}.\end{align} This spectrum is the same as that of graphene, treated within the massless-Dirac-fermion approximation, in a uniform magnetic field~\cite{girvin_yang_2019}. The eigenstates of $h(k)$ are 
\begin{align}
&\psi_n(k)\approx
&\begin{cases}  Q\varphi_0(k)\spin+ & n=0
\vspace{0.2cm}\\ {\displaystyle \frac{ Q}{\sqrt{2}}}\left(\begin{array}{c}{\text{sign}(n)\varphi_{|n|}(k)} \\ {\varphi_{|n|-1}(k)} \end{array}\right) &|n|\geq 1
  \end{cases} \label{effpsi} ,
\end{align}
where 
\begin{align}
    \varphi_n(k)&=\frac{a^{\dag n}}{\sqrt{2^n n!\pi\epsilon}}e^{-k^2/\epsilon}\,
\end{align} are the harmonic-oscillator eigenstates. By using the above effective solutions, we can construct the approximate expression of $\varphi(\theta)$, see Appendix \ref{app:onedangle}. In Fig.~\ref{hsolution}(d), we see that the effective expression $\varphi(\theta)$ is extremely close to the matrix diagonalization result. We show more comparisons for different values of $\epsilon>2$ in Fig.~\ref{varphi} in Appendix \ref{app:onedangle}.

\section{Evolution of Majorana zero modes for a $2\pi$ rotation}\label{sec:Evo}
We now move to study the process of rotating Majorana zero modes on a disk by calculating their dynamical evolution in a uniformly rotating magnetic field. Hence, we focus on the following time-dependent Schr\"odinger equation
\begin{align}
\mathrm{i}\frac{\partial\psi(\theta,t)}{\partial t}&=H_\text{edge}(\theta;\phi(t))\psi(\theta,t)\nonumber\\
&=\omega_0[-\jth \sigma_z-2\epsilon\sin{(\theta-\omega t)}\sigma_y]\psi(\theta,t)\,,\label{pumpeq}
\end{align}
where $H_\text{edge}(\theta;\phi(t))$ is defined in Eq.~(\ref{Hphi}), and the angle $\phi(t)$ rotates uniformly with angular velocity $\omega$. In this case, the evolution operator can be exactly solved as follows (for further details, see Appendix \ref{app:evolution})
\begin{align}
U^\alpha(\phi)=\exp\left(-\mathrm{i}\jth \phi\right)\exp\left(-\mathrm{i}\phi[\alpha h(\theta)-\jth ]\right)\,.\label{eq:Uphi}
\end{align}
Here the parameter $\alpha$ is defined as the ratio of spectral spacing, Eq.~\eqref{eq:Ejpm}, the rotation frequency, $\omega_0/\omega$, and $h(\theta)$ is defined in Eq.~(\ref{htheta}). This parameter controls the adiabaticity of the evolution; the larger $\alpha$ is, the more adiabatic the motion becomes, and $\alpha=0$ corresponds to the sudden approximation. 

\begin{figure}%[tbp]
\centering
\includegraphics[width=0.4\textwidth]{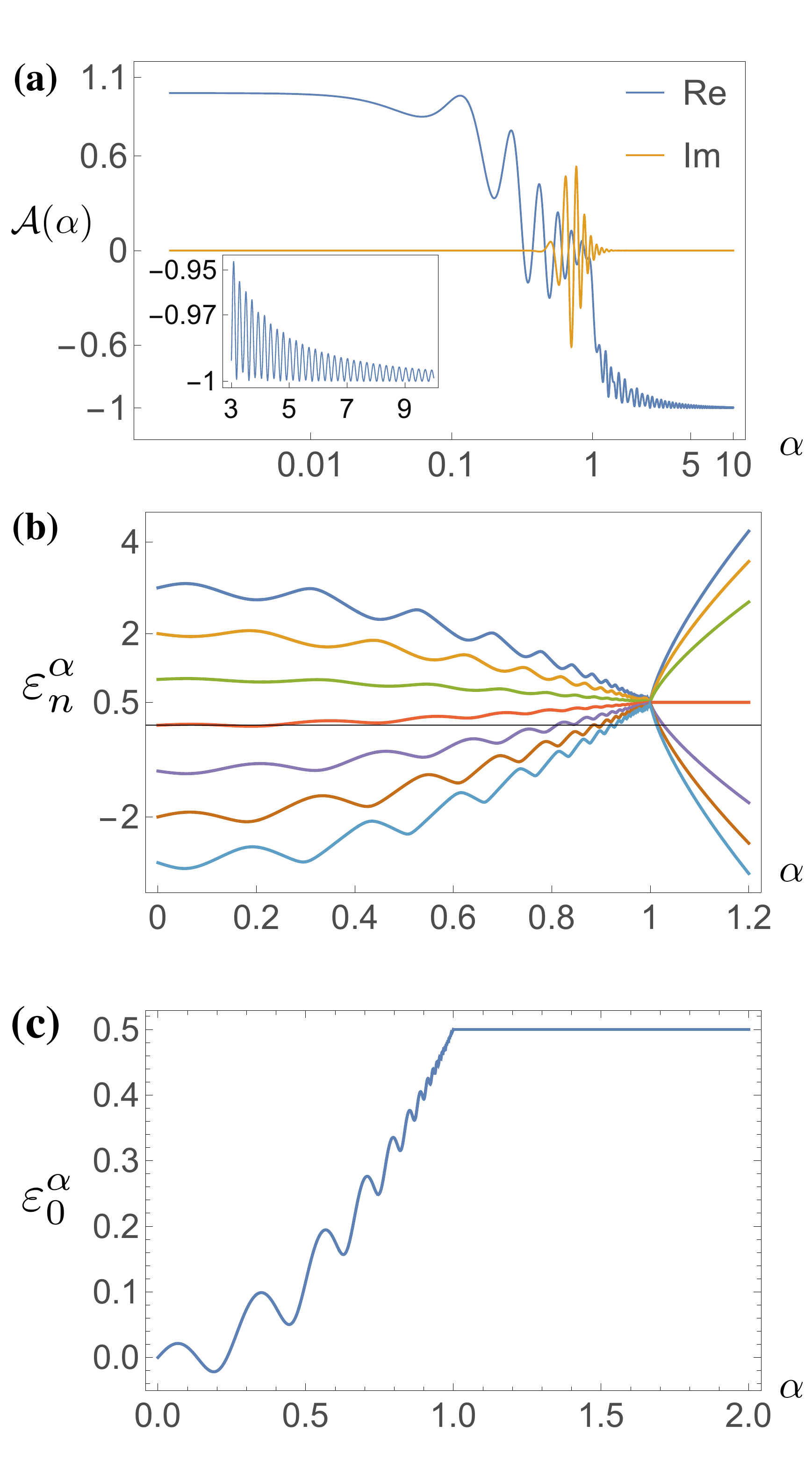}
\caption{(a): The frequency-dependent overlap $\mathcal{A}_0$ between the final state and the initial zero mode. (b)-(c): Quasi-energy spectrum of the evolution operator $\hat{U}^\alpha(2\pi)$ as $\alpha$ changes. There are two distinguishable dynamic regions: a high-frequency region ($\alpha<1$, $\omega>\omega_0$) and a low-frequency region ($\alpha>1$, $\omega<\omega_0$). (b): the quasi-energy spectrum gap closes at $\alpha=1$. (c): A stable $\pi$-mode is present for all $\alpha$ exceeding 1.}\label{evolution_detail}
\end{figure}

One of the ways to analyze the nature of the motion is the overlap between the final state after a $2\pi$ rotation and the initial state, which we shall assume is one of the zero modes.
More specifically, we define the overlap between the final state and the initial zero mode $\psi_0$ within the $\mathbf{J}_1$ subspace as $\mathcal{A}_{0}=\langle\psi_0|\hat{U}^{\alpha}(2 \pi)|\psi_0\rangle$. If we start from a Majorana zero mode, the overlap between the final state and the initial Majorana zero mode can be related to $\mathcal{A}_{0}$ as 
\begin{align}
\langle\psi^m_{{1/2}}|\hat{U}^{\alpha}(2 \pi)| \psi^m_{{1/2}}\rangle&=\text{Re}\{\mathcal{A}_{0}(\alpha) \},\\
\langle\psi^m_{{1}}|\hat{U}^{\alpha}(2 \pi)| \psi^m_{{2}}\rangle&=\text{Im}\{\mathcal{A}_{0}(\alpha) \},\\
\langle\psi^m_2|\hat{U}^{\alpha}(2 \pi)| \psi^m_1\rangle&=-\text{Im}\{\mathcal{A}_{0}(\alpha) \}.
\end{align}
Since $U^\alpha(\phi)$ is generated by $\jth $ and $h(\theta)$ which do not couple the two sets $\mathbf{J}_1$ and $\mathbf{J}_2$, we limit the calculation to $\mathbf{J}_1$. Here we set $\epsilon=5$, $j_{\max}=48+1/2$. We show the resulting overlap $\mathcal{A}_{0}(\alpha)$ in Fig.~\ref{evolution_detail}. For small $\alpha$ ($\omega_0\ll\omega$), we see that $\text{Re}\{\mathcal{A}_0\}$ is very close to 1. This is due to the fact that the system cannot respond quickly enough to the rapid rotation of the magnetic field, so the state has not evolved when the magnetic field comes back to its original value. In the $\alpha>1$ region where $\omega<\omega_0$, $\mathcal{A}_0$ converges to $-1$ gradually as $\alpha$ increases, which corresponds to the $\pi$ phase shift that always appears for particles with half-integer angular momentum after an adiabatic $2\pi$ rotation. Surprisingly, we can see mild oscillations of $\text{Re}\{\mathcal{A}_0\}$ in the $\alpha>1$ region, which reveal non-adiabatic tunneling to excited states. We will discuss this point further in the next section. 

Because of the drastic change in behavior of the overlap we observe when moving from the high-frequency region to the low-frequency region, we see that $\alpha=1$, where $\omega=\omega_0$, is a potential critical point. To understand this potential phase transition of the overlap $\mathcal{A}_{0}(\alpha)$, we now resort to Floquet analysis for the periodically driven topological superconductor and study the spectrum of the one-cycle evolution operator \cite{Floquet1977,floquet1965,Floquet_Majorana_modes}. The Floquet Hamiltonian can be defined through the $2\pi$ rotation evolution operator $U^\alpha(2\pi)$ as
\begin{align}
h^\alpha_F(\theta)=\frac{\mathrm{i}}{2\pi}\ln U^\alpha(2\pi)=\alpha h(\theta)-\jth +\frac{1}{2},
\end{align}
where the factor $1/2$ represents the minus sign introduced by the left exponential in Eq.~\eqref{eq:Uphi}. We now diagonalize the Floquet Hamiltonian and denote the eigenvalues (quasi-energies) as $\varepsilon_n^\alpha$. 

We calculate the Floquet spectrum through a numerical matrix diagonalization within $\mathbf{J}_1 $. To study the behavior of the quasi-energies near $\alpha=1$, we show several curves with quasi-energy close to zero for $\alpha$ ranging from $0$ to $1.2$ in Fig.~\ref{evolution_detail}(b) and also present the $\alpha$-dependence of $\varepsilon_0$ in Fig.~\ref{evolution_detail}(c). As we increase $\alpha$ from $0$, we see that the quasi-energy gap closes at $\alpha=1$ and that a new mode appears at fixed quasi-energy $1/2$, and thus phase $\pi$, in the quasi-energy spectrum when $\alpha$ exceeds $1$. From the Floquet perspective, the collapse of the quasi-energy spectrum gives the critical point for the dynamical phase transition related to the evolution of the zero modes. The pi-phase mode in the low-frequency region (large $\alpha$) provides the mechanism by which the zero mode obtains the adiabatic $\pi$ phase shift for a $2\pi$ rotation. The transition point at $\alpha=1$ gives the upper bound of the frequency by which we can preserve the Majorana zero modes under rotation as $\omega_0/\hbar=\Delta_{\text{exp}}/(\hbar k_FR)$ ($\omega_0$ is defined in Eq.~(\ref{w0})).

By application of the angle translation operator $\exp{i\jth \phi}$ to the wave function, which takes its argument $\theta$ to $\theta+\phi$, we have 
\begin{align}
\mathrm{i}&\frac{\partial\psi(\theta+\phi(t),t)}{\partial t}=\omega\left[h_F^\alpha(\theta)-\frac{1}{2}\right]\psi(\theta+\phi(t),t)\nonumber\\
&=[-\jth \left(\omega I+\omega_0\sigma_z\right)-2\omega_0\epsilon\sin{\theta}\sigma_y]\psi(\theta+\phi(t),t).
\label{comoving}
\end{align}
This equation shows that the Hamiltonian in the co-moving frame is $\omega[h_F^\alpha(\theta)-\frac{1}{2}]$. Thus, the Floquet pi-mode of $h_F$ is the zero mode of this Hamiltonian. For small $\omega$, the solutions to the domain-wall model contain soliton states made up of two counter-propagating chiral states, with the soliton located at the points where the Dirac mass term changes sign. However, when $\omega$ exceeds $\omega_0$, both left-hand and right-hand chiral states all propagate in the same direction, and the domain-wall bound state disappears. In fact, a similar dynamic model has been used to describe moving domain wall solitons in an 
infinite line \cite{soliton,Boosting,Error2019}. In Ref.~\cite{Boosting}, the authors studied the exact spectrum of the boosted model through a global Lorentz transformation in a the flat metric. Unfortunately, the same global transformation fails in our reduced 1D model with circular geometry. Thus, we resort to a Floquet analysis. Note that our analysis can be applied not only to intercalated Bi$_2$Se$_3$ but also to generic Dirac-type models in a disc geometry.

\section{Non-adiabatic corrections to tunnelling in the low-frequency region} \label{sec:nona}
In this section, we use second-order perturbation theory \cite{PhysRevA.55.1653,PhysRevA.58.3439} to interpret the frequency-dependence of the oscillations of $\mathcal{A}_0$ in the low-frequency (large $\alpha$) regime and show that they can be related to non-adiabatic tunneling in the co-rotating frame. 

\begin{figure}[tbp]
\centering
\includegraphics[width=0.44\textwidth]{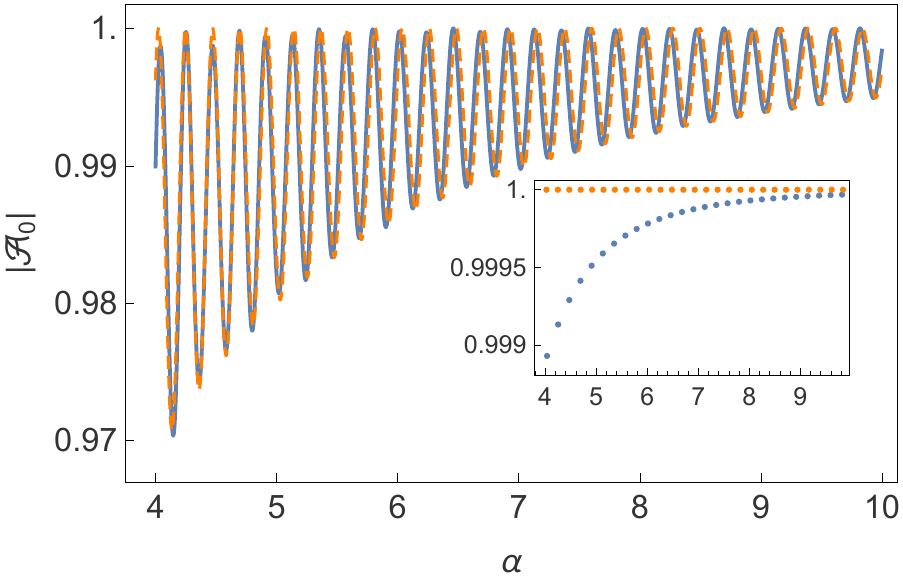}
\caption{A comparison between the exact diagonalization results and the effective results for $|\mathcal{A}_0|$ (see Eq.~\eqref{A0eff_eq}) in the region $\alpha>4$. The blue line is the effective result while the orange dashed-line is the matrix diagonalization result. The insert shows the local maximums in the main figure, which are high-fidelity points with low error.}\label{A0eff}
\end{figure}

To approach this problem, we work in the $k$-representation as set out in Eqs.~(\ref{k_rep}-\ref{eq:Uphi}) and work in a co-moving reference frame, Eq.~(\ref{comoving}). As before, we simplify the algebra by transforming the evolution operator by the operator $Q$, Eq.~(\ref{eq:defQ}) so that the result can be expressed in terms of harmonic oscillator creation and annihilation operators. Using time-dependent perturbation theory, we expand the transformed evolution operator to the second order in $1/\alpha$ and find
\begin{align}
&Q^\dag\exp{-\mathrm{i}\phi[\alpha h(k)-\jth (k)]}Q
\nonumber\\
&\approx\widetilde{U}_{\phi}+\mathrm{i}\int_0^\phi \mathrm{d}\phi_1\widetilde{U}_{\phi-\phi_1} \sqrt{\epsilon}(a+a^\dag)\widetilde{U}_{\phi_1}
\nonumber\\
&-\int_0^\phi \mathrm{d}\phi_1\int_0^{\phi_1}\mathrm{d}\phi_2\widetilde{U}_{\phi-\phi_1} \epsilon(a+a^\dag)\widetilde{U}_{\phi_1-\phi_2}(a+a^\dag)\widetilde{U}_{\phi_2},\label{eq:2ndPT}
\end{align}
where
\begin{align}
&\quad\widetilde{U}_{\phi}=\exp{-\mathrm{i}\alpha\phi h_{\text{eff}}(k)}.
\end{align}
Operating with the above effective evolution operator on the initial zero mode $\psi_0$, we have:
\begin{align}
\exp{-\mathrm{i}\phi[\alpha h(\theta)-\jth ]}\psi_0&\approx\sum\limits_{n=-1}^{1}a_n(\phi)\psi_n,\label{Perturbation}
\end{align}
\begin{align}
a_{0}(\phi) &=1-\frac{\sin{^2(\alpha\epsilon_1\phi/2)}}{2\alpha^2}, \\
 a_{1}(\phi) &=\frac{1-e^{-\mathrm{i} \alpha \epsilon_{1} \phi}}{2 \sqrt{2} \alpha}=a_{-1}^{\ast}(\phi).
\end{align}
In the above equations, $\epsilon_1=\sqrt{4\epsilon}$ is the energy gap between the zero mode and the lowest positive energy state. Since the transition probabilities obey $|a_{0}(\phi) |^2\approx1-(|a_{1}(\phi)|^2 +|a_{-1}(\phi)|^2 )$, the non-adiabatic tunneling in the low-frequency regime mostly happens between $\psi_0$, and $\psi_{\pm1}$ as defined in Eq.~(\ref{effpsi}). Using the result Eq.~(\ref{Perturbation}), we can obtain an effective expression of the overlap $\mathcal{A}_0$ for a full one-cycle rotation, 
\begin{align}
\mathcal{A}_0=-a_{0}(2\pi)=-1+\frac{\sin{^2(\alpha\epsilon_1\pi)}}{2\alpha^2}.\label{A0eff_eq}
\end{align}
This equation accurately describes the behavior of the frequency-dependent oscillation in the overlap $\mathcal{A}_0(\alpha)$. The minus sign originates from the anti-periodic boundary conditions on the wave function. We compare the results of a matrix diagonalization and the effective theory in Fig.~\ref{A0eff}. We can see that the result obtained with the two methods are very close. In the inset of Fig.~\ref{A0eff}, we show the highest and lowest points of the curve $\mathcal{A}_0$. The deviation between exact-diagonalization and effective results in Fig.~\ref{A0eff} is about $0.1\%$. 
From Eq.~(\ref{A0eff_eq}), we can see that the non-adiabatic error can be almost zero when the rotation period of the magnetic field is an integer multiple of the frequency-independent period $2\pi/(\sqrt{4\epsilon}\omega_0)$.

This result can be understood as follows. In the co-rotating frame, the time-dependent evolution is governed by Eq.~(\ref{comoving}). The term $-\jth $ in $h_F$ can make the initial zero mode tunnel to other eigenstates of the equilibrium Hamiltonian. Also, the transition amplitude between the initial zero mode and the final state is equal to $1-\omega^2\sin{^2}\left(\sqrt{4\epsilon}\omega_0t/2\right)/(2\omega_0^2)$, and vanishes when the $t$ is an integer multiple of the tunneling period, i.e., in the perturbative and adiabatic regime (in the case with $\epsilon=5$, the region starts at about $\alpha>4$), at frequencies
\begin{equation} 
\omega_\nu(\Delta\phi)=\frac{\sqrt{4\epsilon}\omega_0\Delta\phi}{2\pi \nu},\end{equation} with $\nu$ an integer, 
the final state after one full rotation of the magnetic field has overlap $-1$ with the initial state. For a $2\pi$ rotation ($\Delta\phi=2\pi$), the integer $\nu$ should be larger than $18$ within the regime $\alpha>4$. Since $\nu$ is large, the difference of the two nearest-neighbor high-fidelity frequencies can be approximated as $\omega_{\nu}(2\pi)-\omega_{\nu+1}(2\pi)\approx\sqrt{4\epsilon}\omega_0/\nu^2$.

\section{Parameter Estimation and Conclusions} \label{sec:conc}

In order to see how realistic our theoretical approach is, we need to estimate the parameters of the model:
From Refs.~\cite{phongPRB,APL2009,experiment_2010,experiment_2011,experiment_2011Hasan,experiment2014}, we estimate $\Delta_{\exp}=0.6\text{meV}$, $k_F=0.1\,\text{\AA}^{-1}$. For a device size $R=10\mu\text{m}$, we determine the dimensionless parameters as $\lambda=10^4$, $\gamma=\Delta_{\exp}^2/(2\bar{\epsilon}_F^2)=4.5\times10^{-6}$, and $\omega_c=\omega_0/\hbar\approx91.2\,\text{MHz}$. To obtain a zero mode, one needs $\epsilon>2$, which in turn implies that the in-plane magnetic field should be larger than $B_0=4\omega_0/\mu_e\approx4\,\text{mT}$. For $\epsilon=5$ as used in the main text, the corresponding magnetic field is about $10\,\text{mT}$. In this case, when the rotation frequency exceeds $\omega_c=91.2\,\text{MHz}$, Majorana pi-modes appear and the region where we can perform quasi-adiabatic 2$\pi$ rotations is $\alpha>4$, corresponding to $\omega<\omega_c/4\approx 22.8\,\text{MHz}$. Since the difference between the two nearest-neighbor high-fidelity frequencies is about $407/\nu^2$MHz and $\nu$ is larger than 18, we require the frequency width of the rotating magnetic field smaller than $0.315$MHz.

In conclusion, we have made progress on two issues: Firstly, we show through a rigorous process how to reduce a 2D Hamiltonian of a second-order $p$-wave superconductor with disc geometry to a 1D Hamiltonian for edge space and build an effective model to give the approximated analytical solution, which improves on the results in Ref.~\cite{phongPRB}, and also allows us to study the evolution in detail; Secondly, we show that the separation between diabatic and adiabatic motion of rotating Majorana zero-modes in the system exhibits an interesting dynamical phase transition at $\alpha=1$. Beyond this transition point, we have been able to give a detailed numerical and analytical analysis of the approach to adiabaticity. Thus the effective matrix model studied here, with its simple solution, could be of further interest. 

We show that at a set of regularly spaced rotation frequencies, we have substantially higher fidelity; by choosing to work at one of these optimal frequencies, one can dramatically reduce quasi-particle poisoning and increase the robustness of the translation of the Majorana states, and thus the manipulation of quantum information in such a device. Our analysis ignores other sources of errors, such as those caused by disorders or fluctuations in the magnetic field, etc. These will be the subject of future work.

\acknowledgements
L.Y. acknowledges funding through China Scholarship Council under Grant 201906230305; A.P. acknowledges support from the European Commission under the EU Horizon 2020 MSCA-RISE-2019 programme (project 873028 HYDROTRONICS) and of the Leverhulme Trust under the grant RPG-2019-363. N.R.W. is supported by STFC grant ST/P004423/1. L.Y. thanks Gabriel Hawkins-Pottier and Hongzheng Zhao for useful discussions.
\appendix 

\section{Symmetry and wave function of edge states}\label{app:wfsym}
The complete expression for the left-hand chiral edge wave function of a $p$-wave superconductor with a disk geometry is \cite{phongPRB}
\begin{align}
&\Psi_{j}^{+}(\rho, \theta)=\mathcal{N}_je^{\mathrm{i} j\theta}\left(\begin{array}{c}
e^{\mathrm{i}\frac{\theta}{2}} f_j(\rho) \\
e^{-\mathrm{i}\frac{\theta}{2}}g_j(\rho) 
\end{array}\right)\otimes\left(\begin{array}{c}
1 \\
0
\end{array}\right)
,\label{wave1}\\
&f_j(\rho)=\text{Im}\{-\sqrt{2\gamma}\kappa_-\kappa_+J_{j+1/2}(\lambda\kappa_-)J_{j+1/2}(\lambda\kappa_+\rho)\},\\
&g_j(\rho)=\text{Im}\{\kappa_-(\kappa_+^2-1-\mathcal{E}_\kappa)J_{j+1/2}(\lambda\kappa_-)J_{j-1/2}(\lambda\kappa_+\rho)\},\\
&\kappa_\pm=(1-\gamma\pm\sqrt{\gamma^2-2\gamma+\mathcal{E}_\kappa^2})^{\frac{1}{2}},\label{keq}\\
&\mathcal{E}_\kappa=\sqrt{(\kappa^2-1)^2+2\gamma\kappa^2},\\
&\mathcal{N}_j=\left(2\pi\int_0^1\rho \mathrm{d}\rho [f_j^2(\rho)+g_j^2(\rho)]\right)^{-\frac{1}{2}}.
\end{align}
The symmetries of the TSC are time-reversal and charge conjugation.
The charge conjugation transformation for the real-space representation is implemented as
\begin{align}
&C=-\tau_x\mathcal{K}\otimes s_z,
\end{align}
where the minus sign in $C$ is for later convenience; the spin matrix $-s_z$ in $C$ adds a minus sign to the $\chi_-$ components in the Nambu spinor. The time-reversal transformation takes the form
\begin{align}
&T=\mathrm{i}\tau_y\mathcal{K}\otimes s_x.
\end{align}
The particle-hole and time-reversal symmetries for the model Hamiltonian without an external field are expressed by the relations
\begin{align}
\{\hat{C},\hat{H}\}&=0,\\
[\hat{T},\hat{H}]&=0,
\end{align}
where $\hat{T}$ and $\hat{C}$ commute with each other.

Combining $T$ and $C$, we can define the chiral operator $S$ as
\begin{align}
&S=TC=\tau_z\otimes \mathrm{i}s_y,\\
&\{\hat{S},\hat{H}\}=0.\label{TPH}
\end{align}
Note that the particle-hole symmetry is conserved with the in-plane Zeeman term.

After applying the in-plane Zeeman field, we break the time-reversal symmetry and thus need to construct new operators to characterize the symmetries of the total Hamiltonian. We define several additional operators as follows
\begin{align}
X&=\tau_0\otimes s_z,\\
O&=TX,\\
\Sigma&=SX=-\tau_z\otimes s_x.\label{TPC}
\end{align}
We then have 
\begin{align}
&[\hat{O},\hat{H}+\hat{H}_Z]=0,\\
&\{\hat{\Sigma},\hat{H}+\hat{H}_Z\}=0.
\end{align}

If we denote the set of chiral edge states  as $|j,\chi\rangle$, satisfying
\begin{align}
&\hat{J}_z|j,\chi\rangle=j|j,\chi\rangle\,,\label{jstate}
\end{align}
and thus 
$j$ is a half-integral angular momentum and $\chi=\pm$ is the chirality. Under the $\hat{T}$, $\hat{C}$, $\hat{O}$ and $\hat{\Sigma}$ transformations, these eigenstates change as follows:
\begin{align}
&\hat{T}|j,\chi\rangle=\chi|-j,-\chi\rangle,\\
&\hat{C}|j,\chi\rangle=|-j,\chi\rangle,\label{Pfg}\\
&\hat{T}\hat{C}|j,\chi\rangle=\chi|j,-\chi\rangle,\\
&\hat{O}|j,\chi\rangle=|-j,-\chi\rangle,\\
&\hat{\Sigma}|j,\chi\rangle=|j,-\chi\rangle.
\end{align}

Using Eq.~(\ref{Pfg}), we find an intertwining relation between the radial functions $f_j(\rho)$ and $g_j(\rho)$,
\begin{align}
\mathcal{N}_jf_j(\rho)=-\mathcal{N}_{-j}g_{-j}(\rho).
\end{align}
 By applying the $\Sigma$ transformation  Eq.~(\ref{TPC}), we see that we can decompose the right-handed  partner for the wave function \eqref{wave1} as
\begin{align}
\Psi_{j}^{-}(\rho, \theta)&=\Sigma\Psi_{j}^{+}(\rho, \theta)\nonumber\\
&=-\tau_z\otimes s_x\mathcal{N}_je^{\mathrm{i} (j-\frac{1}{2})\theta}\left(\begin{array}{c}
e^{\mathrm{i} \theta} f_j(\rho) \\
g_j(\rho) 
\end{array}\right)\otimes\left(\begin{array}{c}
1 \\
0
\end{array}\right)\nonumber\\
&=\mathcal{N}_je^{\mathrm{i} j\theta}\left(\begin{array}{c}
-e^{\mathrm{i}\frac{\theta}{2}} f_j(\rho) \\
e^{-\mathrm{i}\frac{\theta}{2}}g_j(\rho) 
\end{array}\right)\otimes\left(\begin{array}{c}
0 \\
1
\end{array}\right).\label{wave2}
\end{align}

\begin{figure*}[tbp]
\centering\includegraphics[width=1.0\textwidth]{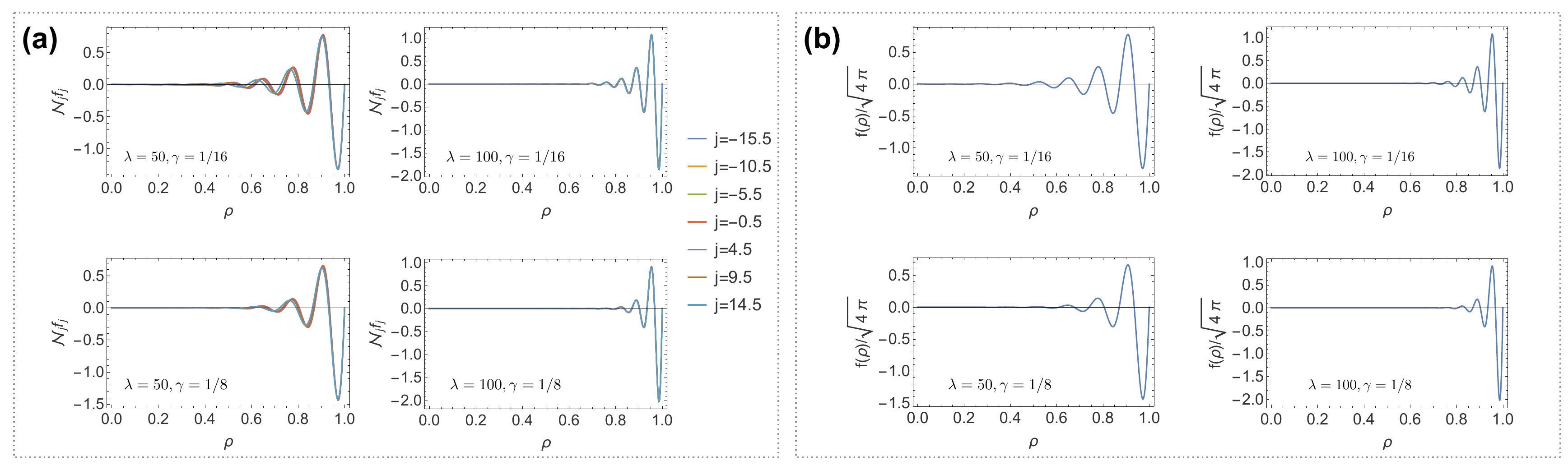}
\caption{Radial wave functions for the chiral edge states $\Psi_j^+$. (a) Exact radial functions with the normalization factor $\mathcal{N}_jf_j(\rho)$ for $-31/2<j<19/2$; (b) Effective radial functions for the same parameters. We see that all radial wave functions are very close near $\rho=1$,  the boundary of the disk, and are very well approximated by a single effective wave function.   }\label{separation_proof}
\end{figure*}

The validity of the reduction of the wave function is based on large $\lambda$ expansion to approximate $f_j(\rho)$ to the first order in $1/\lambda$. Using the asymptotic form of the Bessel function, we have \cite{Arfken1999}
 \begin{align}
 J_{\ell}(z)&\approx\sqrt{\frac{2}{\pi z}}\cos{(z-\frac{\ell\pi}{2}-\frac{\pi}{4})}\nonumber\\
 &=\sqrt{\frac{2}{\pi z}}\frac{1}{2}[e^{\mathrm{i}(z-\frac{\ell\pi}{2}-\frac{\pi}{4})}+e^{-\mathrm{i}(z-\frac{\ell\pi}{2}-\frac{\pi}{4})}],\\
 &\qquad |z|\gg\ell^2-\frac{1}{4}.\nonumber
\end{align}
For large $\lambda$, we have the following approximation:
 \begin{align}
 \kappa_+  &\approx\cos{\xi}+\mathrm{i}\sin{\xi},\\
 \xi&=\frac{1}{2}\arctan{\frac{\sqrt{2\gamma}}{1-\gamma}}.
\end{align}
Since $\gamma$ is small, we know that the real and imaginary parts of $\kappa_+$ are all positive. Thus, we neglect the exponentially decaying part of $ J_{\ell}(z)$:
 \begin{align}
 J_{\ell}(z)\approx\sqrt{\frac{2}{\pi z}}\frac{e^{-\mathrm{i}(z-\frac{\ell\pi}{2}-\frac{\pi}{4})}}{2}.
\end{align}
Using all the approximations mentioned, we obtain the following asymptotic form of $\mathcal{N}_jf_j(\rho)$:
\begin{align}
\mathcal{N}_jf_j(\rho) &\approx\frac{1}{\sqrt{4\pi}}f(\rho)=-\frac{\widetilde{\mathcal{N}}}{\sqrt{4\pi\rho}}e^{\lambda\sin{\xi}\rho}\sin{[\lambda\cos{\xi}(1-\rho)]},\label{radial_fa}\\
\widetilde{\mathcal{N}}&=\left(\int_0^1\mathrm{d}\rho\,e^{2\lambda\sin{\xi}\rho}\sin{^2[\lambda\cos{\xi}(1-\rho)]}\right)^{-\frac{1}{2}}.
\end{align}
In order to apply the asymptotic expansion, $|\lambda\kappa\rho|$ should be very large. Taking $\lambda\gg 1$--in effect equivalent to using a mesoscopic value for $R$ for the small value of $k_F$ estimated in Ref.~\cite{phongPRB})--and taking $\gamma\ll 1$ is sufficient since $\rho\leq 1$ (by definition, $\rho=r/R$) and $\kappa\sim1$ in the above setting. We show the exact results for $\mathcal{N}_jf_j(\rho)$ in Fig.~\ref{separation_proof} (a), and see that these are very similar to the approximate results shown in Fig.~\ref{separation_proof} (b).
Thus, for a general eigenstate wave function, we can reduce it as an unknown angular wave function times the fixed radial one,
 \begin{align}
\Psi(\rho,\theta)&\approx\left\{\psi^+(\theta)\left(\begin{array}{c} \frac{e^{\mathrm{i} \frac{\theta}{2}}}{\sqrt{2}}   \\
- \frac{e^{-\mathrm{i} \frac{\theta}{2}}}{\sqrt{2}} \\0 \\
0
\end{array}\right)-\psi^-(\theta)\left(\begin{array}{c}0 \\
0\\
\frac{e^{\mathrm{i} \frac{\theta}{2}}}{\sqrt{2}}    \\
\frac{e^{-\mathrm{i}\frac{\theta}{2}} }{\sqrt{2}} 
\end{array}\right)\right\}f(\rho),\label{decomposition}\end{align}
where the angular functions $\psi^{\pm}(\theta)$ can be related to the coefficient $c_j^{\pm}$ as
\begin{align}
\psi^+(\theta)&=\frac{1}{\sqrt{2\pi}}\sum_j c_j^+e^{\mathrm{i} j\theta},\\
\psi^-(\theta)&=\frac{1}{\sqrt{2\pi}}\sum_j c_j^-e^{\mathrm{i} j\theta}.
\end{align}
We can combine $\psi^{\pm}(\theta)$ as a spinor field in term of $\theta$, the one-dimensional variable, as \begin{align} \psi(\theta)=\left(\begin{array}{c}\psi^+(\theta)\\ \psi^-(\theta)\end{array}\right).\end{align}
Then the corresponding symmetry operators will be reduced as 
 \begin{align}
 \hat{C}&\to \mathcal{K},\\
 \hat{T}&\to -\mathrm{i}\sigma_y\mathcal{K},\\
 \hat{X} &\to \sigma_z,\\
 \hat{O}&\to \sigma_x\mathcal{K},\\
 \hat{\Sigma}&\to \sigma_x.
\end{align}

%%%%
\section{Tight-binding model calculation}\label{app:TB}
As an additional check for the continuum model used in the main text, we use an effective tight-binding model in this section. This is not based on the estimated experimental parameters but on the same dimensionless parameters used in the main text, which corresponds to a large effective lattice constant to reduce the number of the lattice points of the wave functions.

For simplicity, we work on a square lattice. Firstly, we write down the square-lattice Hamiltonian with left-hand chirality (+1 bulk Chern number) without applying the in-plane magnetic field: 

\begin{align}
  \mathcal{H}_+&=\varepsilon I_{x}\otimes I_{y}\otimes\tau_z-u(m_++m_-+n_++n_-)\otimes\tau_z\nonumber\\&+\mathrm{i}\Delta[(m_+-m_-)\tau_y+(n_+-n_-)\tau_x],\label{spindown_H}\\
      m_{\pm}&=\sum_{m}|m\pm1\rangle\langle m|,\quad n_{\pm}=\sum_{n}|n\pm1\rangle\langle n|,\\
  \varepsilon&=4u-1,\quad u=\frac{1}{(ak_F)^2},\quad   \Delta=\frac{\sqrt{2\gamma}}{2ak_F}.
  \end{align}

Here $a$ is the lattice constant, $(m,n)$ denotes a lattice point with coordinate $(x=ma,y=na)$, and for convenience, the Hamiltonian is scaled by the energy factor $\bar{\epsilon}_F$.
  
\begin{figure*}
\centering\includegraphics[width=0.8\textwidth]{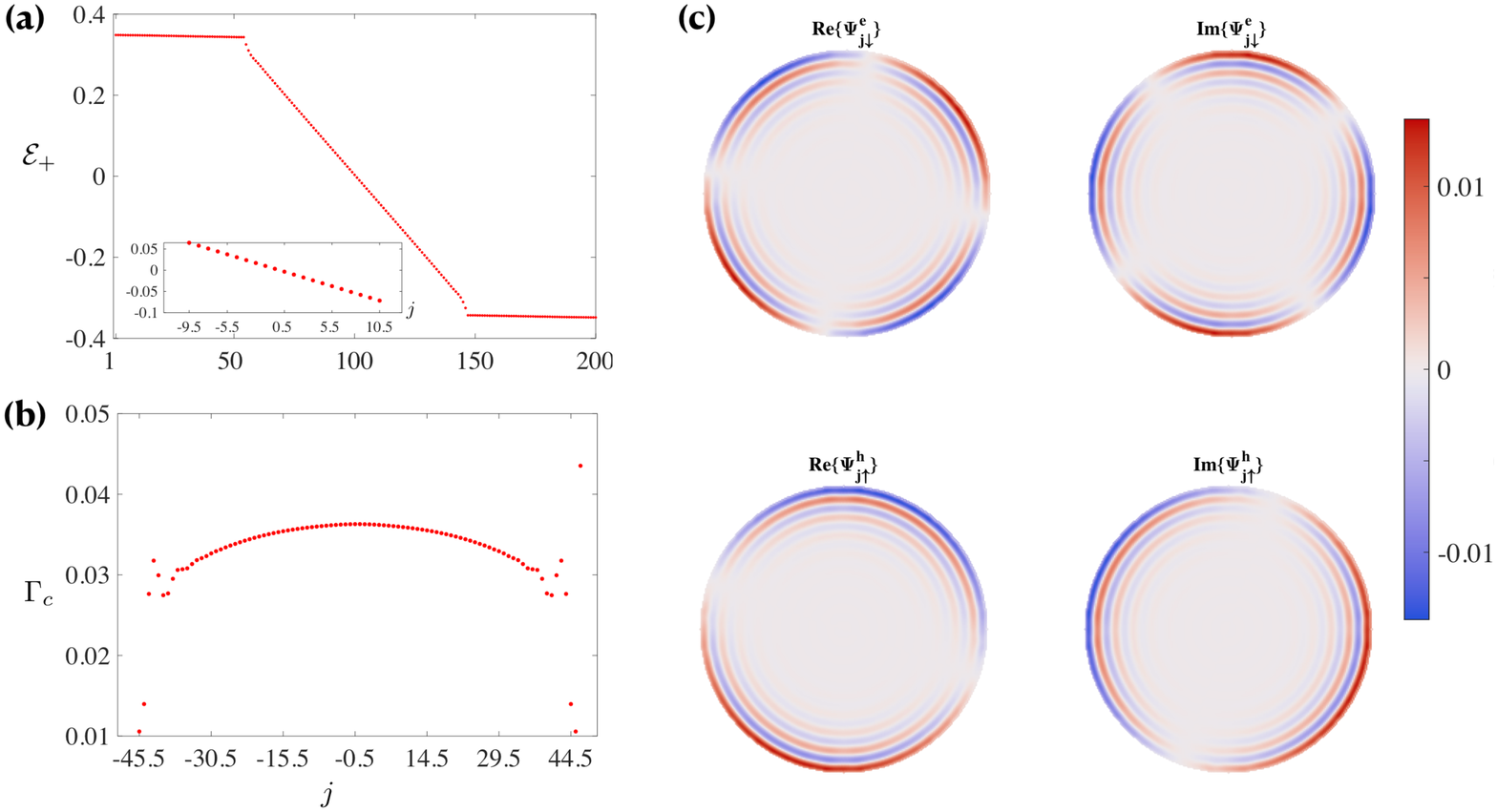}
\caption{Numerical results of the tight-binding model. In (a), we show the energy levels of the left-hand chiral eigenstates and the mid-gap left-hand edge states. The latter is shown in more detail in the inset, where we now label the states by angular momentum. We have checked the relative deviation $\Gamma_c=|\mathcal{E}_{\text{exact}}-\mathcal{E}_{\text{eff}}|/|\mathcal{E}_{\text{eff}}|$ in figure (b). We can see that $\Gamma_c$ is between 1\% and 5\%. We show the two components of the wave function of the $|j=3/2,+\rangle$ edge state in (c). These are either purely real or purely imaginary. We see that the nodal structures in (c) are consistent with the analytical formula Eq.~(\ref{wave1}).\label{lattice_chiral}}
\end{figure*}

In our calculation, we use the following lattice parameters for $\gamma=1/16$:
\begin{align}
  \varepsilon&=15,\quad a=1,\quad u=\frac{1}{k_F^2}=4,\quad \nonumber\\
  \Delta&=\frac{1}{\sqrt{8}}, \quad \lambda=k_FR=50.
  \end{align}
We construct an approximate disk system with radius $R=100$ and solve for the 200 states with energy closet to zero--largely in-gap edge states. We show the related results in Fig.~\ref{lattice_chiral}. In part (a), we can see a clear set of in-gap edge states between the bulk states. As we have discussed in the main text, these edge states can be characterized by the total angular momentum $j$. In the continuum model, we find that the largest allowed angular momentum number $|j_{\max}|$ is the half-integer below $\lambda\sqrt{1-\gamma/2}$. For the case $\lambda=50$, $|j_{\max}|$ should be $48+1/2$. In the Tight-binding calculation, the total number of edge states is about 91, and the corresponding range of $j$ is $[-46+1/2,46+1/2]$. Additionally, we find that the absolute values of the dimensionless energies of bulk states start at $0.34312$, which is very close to the predicted minimum $\sqrt{2\gamma-\gamma^2}\approx0.3480$. We then compare the relative error $\Gamma_c=|\mathcal{E}_{\text{exact}}-\mathcal{E}_{\text{eff}}|/|\mathcal{E}_{\text{eff}}|$ between the numerical results and the effective expression $\mathcal{E}_{\text{eff}}=\mp \sqrt{2 \gamma} j/\lambda $ in (b). We find that this error is between 1\% and 5\%.
Finally, we show the wave function of a left-hand edge state with $j=3/2$ in (c) to check the validity of the expression Eq.~(\ref{wave1}). Recalling that expression, we see that $\Psi_\downarrow^e$ is proportional to $e^{\mathrm{i}(j+1/2)\theta}$ while $\Psi_\uparrow^h$ is proportional to $e^{\mathrm{i}(j-1/2)\theta}$. The wave function we show in (c) clearly has the right nodal structures along the $\theta$ direction determined by $j$.

Based on $\mathcal{H}_+$, we can obtain the time-reversal partner block of it and construct the TSC Hamiltonian with in-plane Zeeman field. 
\begin{align}
  \mathcal{H}(\phi)  &=  \mathcal{H}_+\oplus \mathcal{H}_-+ \mathcal{H}_Z(\phi),\\
  \mathcal{H}_-&=-\mathrm{i}\tau_y\mathcal{H}_+^\ast \mathrm{i}\tau_y\nonumber\\&=-\varepsilon I_{x}\otimes I_{y}\otimes\tau_z+u(m_++m_-+n_++n_-)\otimes\tau_z\nonumber\\&+\mathrm{i}\Delta[(m_+-m_-)\tau_y+(n_+-n_-)\tau_x],\label{TB_lattice}\\
  \mathcal{H}_Z(\phi)&=\mathcal{E}_Z I_{x}\otimes I_{y}\otimes(\cos{(\phi)}\tau_x\otimes s_x-\sin{(\phi)}\tau_y\otimes s_x).
    \end{align}
The components of the complete state are ordered as follows:
\begin{align}
\left(\begin{array}{cccc} |\Psi_\downarrow^e\rangle &
 |\Psi_\uparrow^h\rangle  & -|\Psi_\downarrow^h\rangle  &
|\Psi_\uparrow^e\rangle
\end{array}\right)
\end{align}    

\begin{figure*}
\centering
\includegraphics[width=0.7\textwidth]{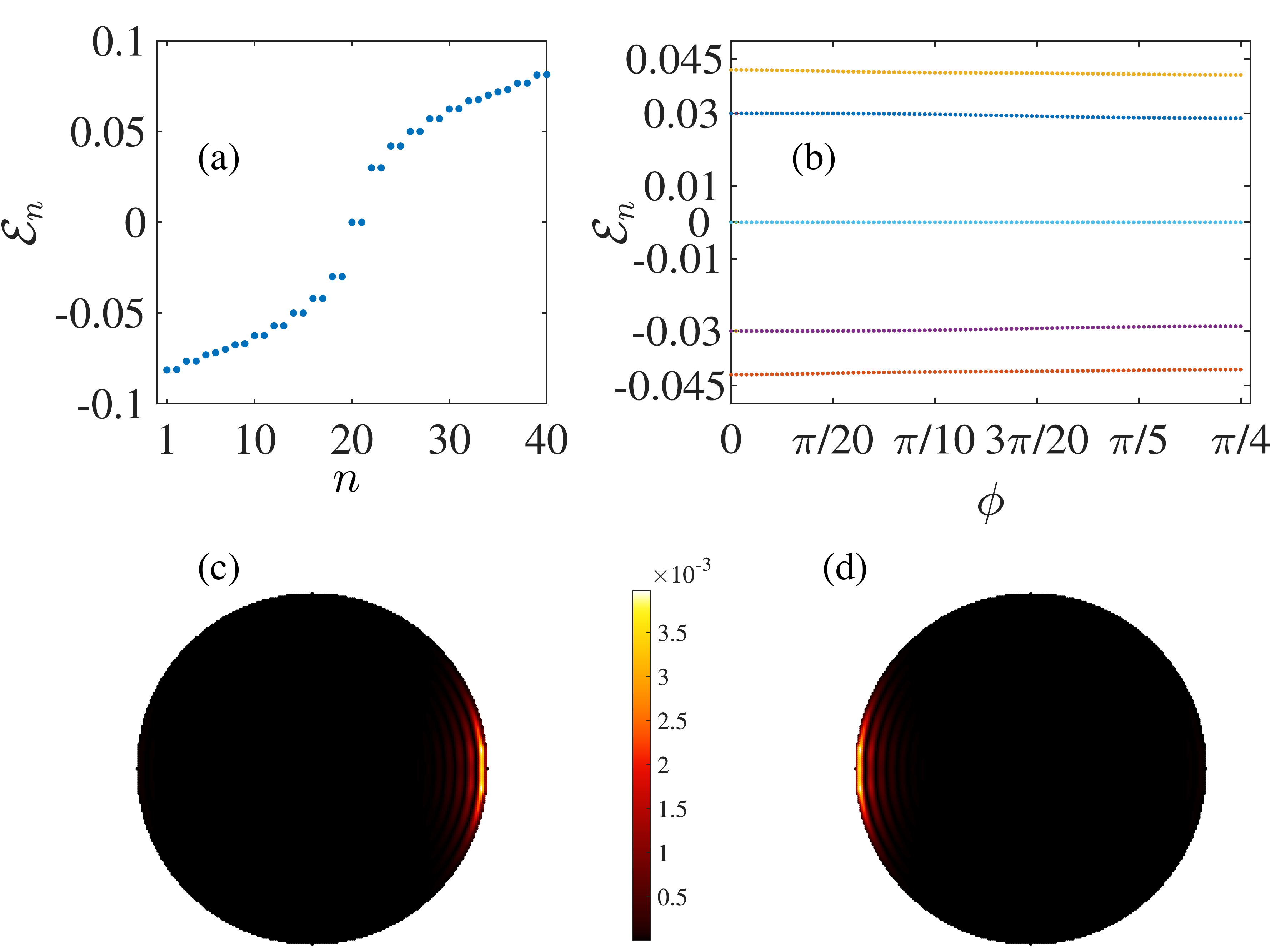}
\caption{Numerical results from the square-lattice model for the topological superconductor under the influence of an in-plane Zeeman field. a): Low-energy spectrum of the square-lattice TSC under in-plane Zeeman field, which is oriented in x-direction; b) Several eigen energies for different orientational angles; c) Probability density of Majorana zero mode $\Psi^m_1$; d) Probability density of Majorana zero mode $\Psi^m_2$.}\label{lattice_z}
\end{figure*}

In order to make a comparison between the tight binding and continuum model possible, we take $\epsilon=5$ which yields $\mathcal{E}_Z=E_Z/\bar{\epsilon}_F=2\epsilon\sqrt{2\gamma}/\lambda=1/(10\sqrt{2})$. The actual numerical magnitude of the eigenenergies for the approximate zero modes is about $10^{-8}$, which is so close to 0 that we can regard them as absolute zero modes. Also, the first and second excited energies from the tight-binding method are about 0.03 and 0.042, respectively while the effective prediction show $\sqrt{2\gamma(4\epsilon)}/\lambda\approx0.0316$ and $\sqrt{2\gamma(8\epsilon)}/\lambda\approx0.0447$ derived from Eq.~(25) in the main text. We show the energy levels around zero energy in Fig.~\ref{lattice_z} (a), where two zero modes localize in the middle of the spectrum. For the square-lattice Tight-binding calculation with a limited number of lattice points, the edge of the disk is not a perfect circle. Therefore, we see the energy levels deviate slightly as $\phi$ increases in the region $\phi\in[0,\pi/4]$ in Fig.~\ref{lattice_z} (b). However, the deviation is small enough for us to ignore. 

Lastly, we draw the probability density of the two Majorana zero modes when $\phi=0$ in Figs.~\ref{lattice_z} (c) and (d). The probability density is defined as 
 \begin{align}
 \rho(\vec{r})=|\Psi_\downarrow^e(\vec{r})|^2+|\Psi_\downarrow^h(\vec{r})|^2+|\Psi_\uparrow^e(\vec{r})|^2+|\Psi_\uparrow^h(\vec{r})|^2.
 \end{align}    
 %%%%
\section{Angular wave function of Majorana zero mode}\label{app:onedangle}
In this appendix, we give the expressions of the angular wave functions of the two Majorana zero modes. We can expand the zero-mode wave functions in the two subspaces $\mathbf{J}_1$ and $\mathbf{J}_2$ as 
\begin{align}
\psi_0(\theta)&=\sum\limits_k c^+_{2k-\frac{1}{2}} e^{\mathrm{i}(2k-1/2)\theta} |+\rangle+c^-_{2k+\frac{1}{2}} e^{\mathrm{i}(2k+1/2)\theta} |-\rangle,\\
\psi^\ast_0(\theta)&=\sum\limits_k c^+_{2k-\frac{1}{2}} e^{-\mathrm{i}(2k-1/2)\theta} |+\rangle+c^-_{2k+\frac{1}{2}} e^{-\mathrm{i}(2k+1/2)\theta} |-\rangle,
\end{align}
where $k$ is an integer. Translating $\theta$ to $\theta+\pi$, we see that they obey the following relations
\begin{align}
\psi_0(\theta+\pi)&=\mathrm{i}\,\sigma_z\psi_0(\theta)\label{psi0_sum}\\
\psi^\ast_0(\theta+\pi)&=-\mathrm{i}\,\sigma_z\psi^\ast_0(\theta).
\end{align}
According to the definition given in Eq.~(\ref{eq:main_psi_1_m}) and (\ref{eq:main_psi_2_m}) of the Majorana zero modes, we have 
\begin{align}
&\psi_{1}^{m}(\theta+\pi)=-\sigma_{z} \psi_{2}^{m}(\theta), \\
&\psi_{2}^{m}(\theta+\pi)=\sigma_{z}\psi_{1}^{m}(\theta).\label{Majorana_mutual}
\end{align}
Applying Eq.~(\ref{psi_symmetry}), we can express them as
\begin{align}
\psi^m_{1}(\theta)&=\frac{\psi_0^+(\theta)+\psi_0^-(\theta)}{\sqrt{2}}\left(\begin{array}{c}1\\1
\end{array}\right)\nonumber\\
&=\frac{2\text{Re}\{\psi_0^+(\theta)\}}{\sqrt{2}}\left(\begin{array}{c}1\\1
\end{array}\right)\nonumber\\
&=\varphi(\theta)\frac{|+\rangle+|-\rangle}{\sqrt{2}},\\
\psi^m_{2}(\theta)&=\sigma_{z}\psi_{1}^{m}(\theta-\pi)\nonumber\\
&=\varphi(\theta-\pi)\frac{|+\rangle-|-\rangle}{\sqrt{2}}.
\end{align}
Note that $\varphi(\theta)$ is an even function.

\begin{figure*}
\centering\includegraphics[width=0.8\textwidth]{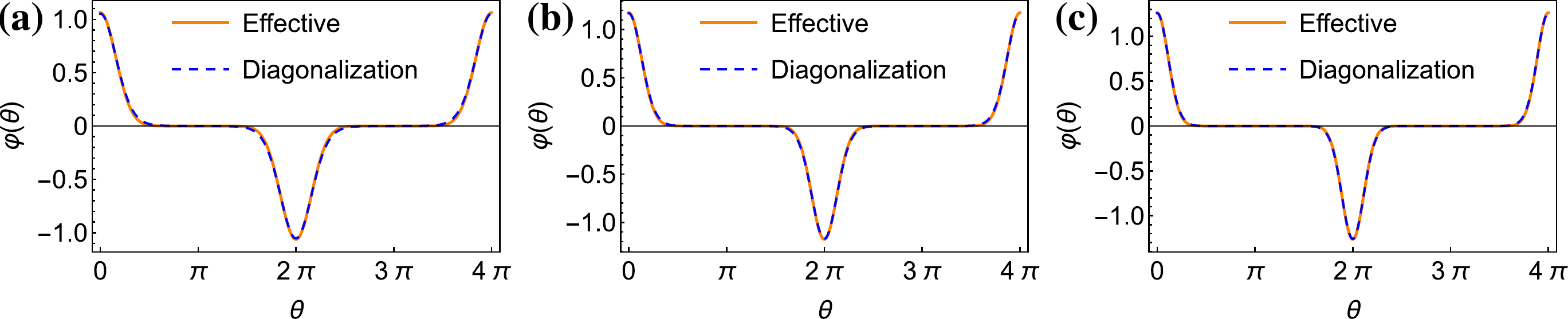}
\caption{The angular function $\varphi(\theta)$ of the Majorana zero mode when $\epsilon=2$ (a), $\epsilon=3$ (b) and $\epsilon=4$ (c). We can see that the effective results are close to the matrix diagonalization results for these cases. Note that the zero modes appear after $\epsilon$ exceeds 2.} \label{varphi}
\end{figure*}
We find that the one-dimensional reduced wave functions of the two Majorana zero modes can be expressed by Jacobi Theta functions\cite{theta_function}:
\begin{align}
\psi^m_1(\theta)&=\mathscr{N}_\epsilon\Theta_2(\frac{\theta}{2},e^{-\frac{1}{4\epsilon}})\frac{|+\rangle+|-\rangle}{\sqrt{2}},\\
\psi^m_2(\theta)&=\mathscr{N}_\epsilon\Theta_2(\frac{\theta-\pi}{2},e^{-\frac{1}{4\epsilon}})\frac{|+\rangle-|-\rangle}{\sqrt{2}}\nonumber\\
&=\mathscr{N}_\epsilon\Theta_1(\frac{\theta}{2},e^{-\frac{1}{4\epsilon}})\frac{-|+\rangle+|-\rangle}{\sqrt{2}}.
\end{align}
The $\Theta$ functions are defined as follows:
\begin{align}
\Theta_1(z,q)&=\sum_{n=-\infty}^{\infty}q^{n+\frac{1}{2}}e^{\mathrm{i}(2n+1)z},\\
\Theta_2(z,q)&=\sum_{n=-\infty}^{\infty}(-1)^{n}q^{n+\frac{1}{2}}e^{\mathrm{i}(2n+1)z}.
\end{align}
Note that $\Theta_1(z,q)$ is an odd function of $z$ while $\Theta_2(z,q)$ is even. We have drawn several curves of $\varphi(\theta)=\mathscr{N}_\epsilon\Theta_2(\theta/2,e^{-1/(4\epsilon)})$ for the cases $\epsilon=2,3,4$ in Fig.~(\ref{varphi}).

When neglecting the boundary condition $\psi(\theta+2\pi)=-\psi(\theta)$, we can linearize the Dirac mass term ($\propto \sin{\theta}$) in $h(\theta)$ (defined in Eq.~(\ref{htheta})). This corresponds to placing one Majorana zero mode around $\theta=0$ while placing the other one at infinity. Then the Hamiltonian can be written as
\begin{align}
\label{app:c_h_theta_def}
h(\theta)&\approx-p_\theta\sigma_z-2\epsilon\theta\sigma_y\nonumber\\
&=2\sqrt{\epsilon}[\mathrm{i}\frac{\tilde{a}^\dag}{2}\left(\begin{array}{c} {1}\\{1}
\end{array}\right)\left(\begin{array}{cc} {-1}&{1}
\end{array}\right)-\mathrm{i}\frac{\tilde{a}}{2}\left(\begin{array}{c} {-1}\\{1}
\end{array}\right)\left(\begin{array}{cc} {1}&{1}
\end{array}\right)],
\end{align}
where
\begin{align}
p_\theta&=\mathrm{i}(\tilde{a}^\dag-\tilde{a})\sqrt{\epsilon},
\end{align}
and
\begin{align}
\theta&=\frac{\tilde{a}^\dag+\tilde{a}}{2\sqrt{\epsilon}}.
\end{align}
The above equations define the operator $\tilde{a}$ as $\tilde{a} = \sqrt{\epsilon}\theta+\mathrm{i}p_\theta/(2\sqrt{\epsilon})$, which satisfies $[\tilde{a},\tilde{a}^\dag]=1$. By applying this om the general symmetric wave function $\varphi(\theta)(|+\rangle+|-\rangle)$, we find a zero energy state when $\tilde{a}\varphi(\theta)=0$ and this gives the solution 
\begin{align}
\varphi(\theta)\propto e^{-\epsilon\theta^2}
\end{align}
Thus we see $\varphi(\theta)$ is a Gaussian function with width $1/\sqrt{2\epsilon}$. Hence,
the ground state of the Hamiltonian~(\ref{app:c_h_theta_def}), {\it i.e.} the Majorana zero mode, can be expressed as:
\begin{align}
\psi^m_1(\theta)\propto e^{-\epsilon\theta^2}\frac{|+\rangle+|-\rangle}{\sqrt{2}}.
\end{align}
 The energies of the two Majorana modes are exactly zero when the two are far away from each other and do not have any overlap. Using Eq.~(\ref{Majorana_mutual}), we can find that the second Majorana zero mode is located around $\theta=\pi$ and has the form
\begin{align}
\psi^m_2(\theta)\propto e^{-\epsilon(\theta-\pi)^2}\frac{|+\rangle-|-\rangle}{\sqrt{2}}.
\end{align}
If we look at a system with both Majorana zero modes present (at $\theta=0$ and $\theta=\pi$),  the overlap between these two can give a tunneling splitting, thus adding to the energy. Therefore, to suppress this splitting, the angular width of the two Majorana should be small enough corresponding to large $\epsilon$.

\section{Evolution of the Majorana quasi-particle}\label{app:evolution}
\subsection{Evolution of the Bogoliubov-de-Gennes quasi-particle}
The Heisenberg equation for the quasi-particle of our driven TSC in the Fock representation is
\begin{align}
\frac{d\gamma_{\text{BdG}}^\dag(t)}{dt}=-\mathrm{i}[\hat{H}_{\text{sc}}(\phi(t)),\gamma_{\text{BdG}}^\dag(t)].
\end{align}
The Hamiltonian $\hat{H}_{\text{sc}}(\phi)$ is constructed by $\mathcal{H}(\phi)$ in Eq.~(\ref{TB_lattice}):
\begin{align}
\hat{H}_{\text{sc}}(\phi)&=\frac{\bar{\epsilon}_F}{2}\mathbf{C}^\dag \mathcal{H}(\phi)\mathbf{C}=\frac{1}{2}H_{\mathbf{r}_1\mathbf{r}_2\tau_1\tau_2}^{\sigma_1\sigma_2}(\phi)c_{\mathbf{r}_1\sigma_1}^{-\tau_1}c_{\mathbf{r}_2\sigma_2}^{\tau_2},\\
       c_{\mathbf{r}\sigma}^{\tau}&=\left\{\begin{array}{cc}
c_{\mathbf{r}\sigma} &\tau=1 \\c_{\mathbf{r}\sigma}^{\dag}& \tau=-1 \end{array}\right..  
\end{align}
Using the PHS and the hermiticity, $H_{\mathbf{r}_1\mathbf{r}_2-\tau_1-\tau_2}^{\sigma_1\sigma_2}=H_{\mathbf{r}_1\mathbf{r}_2\tau_1\tau_2}^{\sigma_1\sigma_2\ast}=H_{\mathbf{r}_2\mathbf{r}_1\tau_2\tau_1}^{\sigma_2\sigma_1}$, and the anti-commutation relation of electrons operators, $\{ c_{\mathbf{r}\sigma}^{\tau}, c_{\mathbf{r}^{\prime}\sigma^{\prime}}^{\tau^{\prime}}\}=\delta_{\mathbf{r},\mathbf{r}^{\prime}}\delta_{\sigma,\sigma^{\prime}}\delta_{-\tau,\tau^{\prime}}$, we obtain the following equation:
\begin{align}
\mathrm{i}\frac{\partial\langle\mathbf{r}_1,\tau_1,\sigma_1|\psi(t)\rangle}{\partial t}&=H_{\mathbf{r}_1\mathbf{r}_2\tau_1\tau_2}^{\sigma_1\sigma_2}\langle\mathbf{r}_2,\tau_2,\sigma_2|\psi(t)\rangle,\\
\mathrm{i}\frac{\partial|\psi(t)\rangle}{\partial t}&=|\mathbf{r}_1,\tau_1,\sigma_1\rangle H_{\mathbf{r}_1\mathbf{r}_2\tau_1\tau_2}^{\sigma_1\sigma_2}\langle\mathbf{r}_2,\tau_2,\sigma_2|\psi(t)\rangle.\label{evolution_fock}
\end{align}
If we define 
\begin{align}
H=|\mathbf{r}_1,\tau_1,\sigma_1\rangle H_{\mathbf{r}_1\mathbf{r}_2\tau_1\tau_2}^{\sigma_1\sigma_2}\langle\mathbf{r}_2,\tau_2,\sigma_2|,
\end{align} 
then we can have the time-dependent Schrodinger equation for the quasi-particle wave function, which is equivalent to the Heisenberg equation:
\begin{align}
\mathrm{i}\frac{d|\Psi(t)\rangle}{dt}&=H(\phi(t))|\Psi(t)\rangle,\label{evolution_abstract}\\
\dot{\phi}(t)&=\omega(t).
\end{align}

\subsection{Derivation of the Evolution operator in both approaches}
To derive an analytical expression for the evolution operator, we first rewrite the Zeeman term as
\begin{align}
H_Z(\phi)&=E_Z\mathcal{G}(\phi)\tau_x\otimes s_x\mathcal{G}^{\dag}(\phi),\\
\mathcal{G}(\phi)&=\begin{smallmatrix}\left(\begin{array}{cc}{e^{\mathrm{i}\phi/2}} & {0} \\ {0}&{e^{-\mathrm{i}\phi/2}}\end{array}\right)\end{smallmatrix}\otimes s_0.\label{eq:Gphi}
\end{align}
Remember that the Hamiltonian for the disk system in long-wave approximation can be written as
\begin{align}
&\mathcal{G}^{\dag}(\phi)H(\mathbf{r},\phi)\mathcal{G}(\phi)=H(\mathbf{r}^{\prime},0),\\
&\mathbf{r}^{\prime}=\mathcal{R}_{\text{2D}}(\phi)\mathbf{r}=r(\cos{(\theta-\phi}),\sin{(\theta-\phi})).
\end{align}
Here $\mathcal{R}_{\text{2D}}(\phi)$ is the operator that implements a two-dimensional clockwise rotation over an angle $\phi$. 
We now apply the transformation $\mathcal{G}^{\dag}(\phi)$ \eqref{eq:Gphi}  to the wave function to obtain the time-dependent equation in a co-moving framework,
\begin{align}
\mathrm{i}\frac{\partial\Psi^{\prime}(\mathbf{r}^{\prime}(\mathbf{r},t),t)}{\partial t}&=[H(\mathbf{r}^{\prime},0)-\mathrm{i}\mathcal{G}^{\dag}(\phi)\frac{\partial\mathcal{G}(\phi)}{\partial t}]\Psi^{\prime}(\mathbf{r}^{\prime},t)\nonumber\\
&=[H(\mathbf{r}^{\prime},0)+\frac{\omega(t)}{2}\tau_z\otimes s_0]\Psi^{\prime}(\mathbf{r}^{\prime},t),\label{eq:rotframe}
\end{align}
and
\begin{align}
\Psi^{\prime}(r,\theta-\phi,t)&=\mathcal{G}^{\dag}(\phi)\Psi(r,\theta,t).
\end{align}
We see that
\begin{align}
\frac{\partial\Psi^{\prime}(\mathbf{r}^{\prime}(\mathbf{r},t),t)}{\partial t}&=\frac{\partial\Psi^{\prime}(r,\theta,t)}{\partial t}-\omega(t)\frac{\partial\Psi^{\prime}(r,\theta,t)}{\partial\theta},
\end{align}
\begin{align}
&\mathrm{i}\frac{\partial\Psi^{\prime}(r,\theta^{\prime},t)}{\partial t}=\nonumber\\&[H(r,\theta^{\prime},0)+\frac{\omega(t)}{2}\tau_z\otimes s_0+\mathrm{i}\omega(t)\frac{\partial}{\partial\theta^{\prime}}]\Psi^{\prime}(r,\theta^{\prime},t),\label{eqa:fR}
\end{align}
which can be summarised as
\begin{align}
\mathrm{i}\frac{\partial\Psi^{\prime}(r,\theta,t)}{\partial t}&=[H(r,\theta,0)-\omega(t) J_z(\theta)]\Psi^{\prime}(r,\theta,t).\label{eqb:fR}
\end{align}
In this equation, $J_z(\theta)=(-\mathrm{i}\partial_\theta-\frac{1}{2}\tau_z)\otimes s_0$, see Eq.~\ref{eq:Jz}. For a constant rotation frequency $\omega(t)=\omega$,
the operator $H(r,\theta,0)-\omega J_z(\theta)$ on the right-hand side of Eq.~(\ref{eqb:fR}) is independent on $t$, so we can directly integrate the equation and obtain the evolution operator %for $\Psi^{\prime}(r,\theta,t)$:
\begin{align}
U^{\prime}(\mathbf{r},t)&=\exp{\{-\mathrm{i}t[H(r,\theta,0)-\omega J_z(\theta)]\}}\,.
\end{align}
Thus, $\Psi(r,\theta,t)$ becomes
\begin{align}
\Psi(r,\theta,t)&=\mathcal{G}(\phi)\psi^{\prime}(r,\theta-\phi,t)\nonumber\\
&=e^{-\partial_\theta \phi}\mathcal{G}(\phi)U^{\prime}(\phi)\Psi(r,\theta,0).
\end{align}
Therefore, the one-period evolution operator is:
\begin{align}
U_{2\pi}(\mathbf{r})&=\exp\left\{\mathrm{i}2\pi\left[-\alpha h(\mathbf{r})+J_z(\theta)-\frac{1}{2}\right]\right\},\label{2dU}\\
h(\mathbf{r})&=\frac{\lambda}{\sqrt{2\gamma}}\mathcal{H}(\mathbf{r},\phi=0).
\end{align}
Using the decomposition in Eq.~(\ref{decomposition}), we can approximately reduce this operator from 2D to 1D, and we see the one-period evolution is described by the reduced Floquet Hamiltonian
\begin{align}
h_F=\alpha h(\mathbf{\theta})-p_\theta(\theta)+\frac{1}{2}.
\end{align}

We can also derive the one-dimensional evolution operator through the reduced equation:
\begin{align}
\mathrm{i}\frac{\partial\psi(\theta,t)}{\partial t}&=\omega_0[-p_\theta\sigma_z-2\epsilon\sin{(\theta-\omega t)}\sigma_y]\psi(\theta,t).
\end{align}
After an unitary transformation $\exp{(ip_\theta\phi)}$ to translate $\theta\to\theta+\phi$, we have:
\begin{align}
\mathrm{i}\frac{\partial\psi^{\prime}(\theta,\phi)}{\partial\phi}&=[\alpha h(\theta)-p_\theta]\psi^{\prime}(\theta,\phi),\\
\psi^{\prime}(\theta,\phi)&=\exp{(\mathrm{i}p_\theta\phi)}\psi(\theta,\phi).
\end{align}
This evolution equation is actually equivalent to Eq.~(\ref{comoving}) in the main text. From this, we can determine the evolution operator as
\begin{align}
U(\phi)&=\exp{[-\mathrm{i}(p_\theta-\frac{1}{2})\phi]}\exp{[-\mathrm{i}\phi h_F(\theta)]},\\
h_F(\theta)&=\alpha h(\mathbf{\theta})-p_\theta(\theta)+\frac{1}{2}.
\end{align}

\bibliography{majorana}

\end{document}